\documentclass[journal]{vgtc}                

  \pdfoutput=1\relax                   
  \usepackage{graphicx}                
  \DeclareGraphicsExtensions{.pdf,.png,.jpg,.jpeg} 

\graphicspath{{figures/}{pictures/}{images/}{./}} 

\usepackage{microtype}                 
\PassOptionsToPackage{warn}{textcomp}  
\usepackage{textcomp}                  
\usepackage{mathptmx}                  
\usepackage{times}                     
\usepackage{cite}                      
\usepackage{tabu}                      
\usepackage{booktabs}                  
\usepackage{wrapfig}
\usepackage{xcolor}
\usepackage{subfigure}
\usepackage{colortbl}
\usepackage{multirow}
\usepackage{amssymb}
\usepackage{amsmath}
\usepackage{xspace}
\usepackage{balance}
\usepackage{colortbl}
\usepackage{url}

\definecolor{color_A}{HTML}{E31A1C}
\definecolor{color_B}{HTML}{FB9A99}
\definecolor{color_C}{HTML}{1F78B4}
\definecolor{color_D}{HTML}{A6CEE3}
\definecolor{color_DD}{HTML}{FEFDE1}

\usepackage{xparse}
\usepackage{tikz}
\usetikzlibrary{matrix,backgrounds, calc}
\pgfdeclarelayer{myback}
\pgfsetlayers{myback,background,main}

\tikzset{mycolor/.style = {line width=1bp,color=#1}}%
\tikzset{myfillcolor/.style = {draw,fill=#1}}%

\NewDocumentCommand{\highlight}{O{blue!40} m m}{%
\draw[mycolor=#1] (#2.north west)rectangle (#3.south east);
}

\NewDocumentCommand{\fhighlight}{O{blue!40} m m}{%
\draw[myfillcolor=#1] ([shift={(-0.05,0.05)}]#2.north west)rectangle ([shift={(0.05,-0.05)}]#3.south east);
}

\NewDocumentCommand{\dhighlight}{O{blue!40} m m}{%
\draw[myfillcolor=#1] ([shift={(-0.1,0.1)}]#2.north west)rectangle ([shift={(0.1,-0.1)}] #3.south east);
}

\usetikzlibrary{matrix}
\DeclareMathOperator{\rank}{rank}

\newcommand{\myparagraph}[1]{\smallskip\noindent\textbf{#1}\space}

\newcommand{\ie}{\emph{i.e.}\xspace}
\newcommand{\eg}{\emph{e.g.}\xspace}

\newcommand{\ck}{\checkmark}

\newcommand{\URL}[1]{{\scriptsize{\url{#1}}}}
\newcommand{\urlfootnote}[1]{\footnote{\footnotesize{\url{#1}}}}

\newcommand{\squishlist}{
   \begin{list}{$\bullet$}
    { \setlength{\itemsep}{0pt}      \setlength{\parsep}{3pt}
      \setlength{\topsep}{3pt}       \setlength{\partopsep}{0pt}
      \setlength{\leftmargin}{1.5em} \setlength{\labelwidth}{1em}
      \setlength{\labelsep}{0.5em} } }

\newcommand{\squishlisttwo}{
   \begin{list}{$\bullet$}
    { \setlength{\itemsep}{0pt}    \setlength{\parsep}{0pt}
      \setlength{\topsep}{0pt}     \setlength{\partopsep}{0pt}
      \setlength{\leftmargin}{2em} \setlength{\labelwidth}{1.5em}
      \setlength{\labelsep}{0.5em} } }

\newcommand{\squishend}{
    \end{list}  }

\onlineid{0}

\vgtccategory{Research}
\vgtcpapertype{application}

\title{CrimAnalyzer: Understanding Crime Patterns in S\~ao Paulo City}

\author{Germain Garcia-Zanabria, Jaqueline Silveira, Jorge Poco~\textit{Member, IEEE}, Afonso Paiva, Marcelo~Batista~Nery,\\ Claudio~T.~Silva~\textit{Member, IEEE}, Sergio Adorno, Luis Gustavo Nonato~\textit{Member, IEEE}}
\authorfooter{
\item
    Germain Garcia-Zanabria, Jaqueline Silveira, and Afonso Paiva are with ICMC-USP, S\~ao Carlos, Brazil. E-mail: \{germaingarcia,alva.jaque\}@usp.br, apneto@icmc.usp.br
\item
 Marcelo Batista Nery is with RIDC -FAPESP and  Institute of Advanced Studies – Global Cities Program. E-mail: mbnery@gmail.com
\item Sergio Adorno is with NEV-USP, S\~ao Paulo, Brazil. E-mail: marsadorno@usp.br
\item Jorge Poco is with Fundação Get\'ulio Vargas, Brazil and Universidad Cat\'olica San Pablo. E-mail: {jorge.poco@fgv.br}
\item
   Claudio Silva is with New York University, USA. E-mail: {csilva@nyu.edu} 
\item
    Luis Gustavo Nonato is with ICMC-USP, S\~ao Carlos, Brazil and New York University, USA. E-mail: {gnonato@icmc.usp.br}
}

\shortauthortitle{Garcia-Zanabria \MakeLowercase{\textit{et al.}}: CrimAnalyzer}

\abstract{
S\~{a}o Paulo is the largest city in South America, with criminality rates as large as the city. The number and type of crimes varies considerably around the city, assuming different patterns 
depending on urban and social characteristics of each particular location. 
Previous works have mostly focused on the analysis of crimes with the
intent of uncovering patterns associated to social factors, time seasonality, and urban routine activities.
Therefore, those studies and tools are more global in the sense that they are not designed to investigate specific regions
of the city such as particular neighborhoods, avenues or public areas. 
Enabling tools able to explore particular locations of the city is essential for domain experts to accomplish their analysis in a bottom-up fashion, 
more clearly revealing how local urban features related to mobility, passersby behavior, and presence of public infrastructures 
such as terminals of public transportation and schools, can influence the quantity and type of crimes. 
In this paper we present CrimAnalyzer, a visualization assisted analytic tool that allows users to analyze the behavior of crimes 
in specific regions of a city. 
The system allows users to identify local hotspots, their pattern of crimes, and  how the hotspots and corresponding crime patterns 
change over time.
CrimAnalyzer has been developed from the demand of a team of experts in criminology and
deals with three major challenges in this context, $i)$ flexibility to explore local regions and understand their crime patterns,
$ii)$ identification of spatial crime hotspots that might not be the most prevalent ones in terms of the number of crimes but are important
enough to be investigated, and 
$iii)$ understand the dynamic of crime patterns and types over time.
The effectiveness and usefulness of the proposed system are demonstrated by qualitative and quantitative comparisons as well as case studies involving real data and run by domain experts.
The experiments show the capability of CrimAnalyzer in identifying crime-related phenomena.

} 
\keywords{Crime Data, Spatio-Temporal Data,  Visual Analytics, Non-Negative Matrix Factorization}

\teaser{
  \centering
  \includegraphics[width=.84\linewidth]{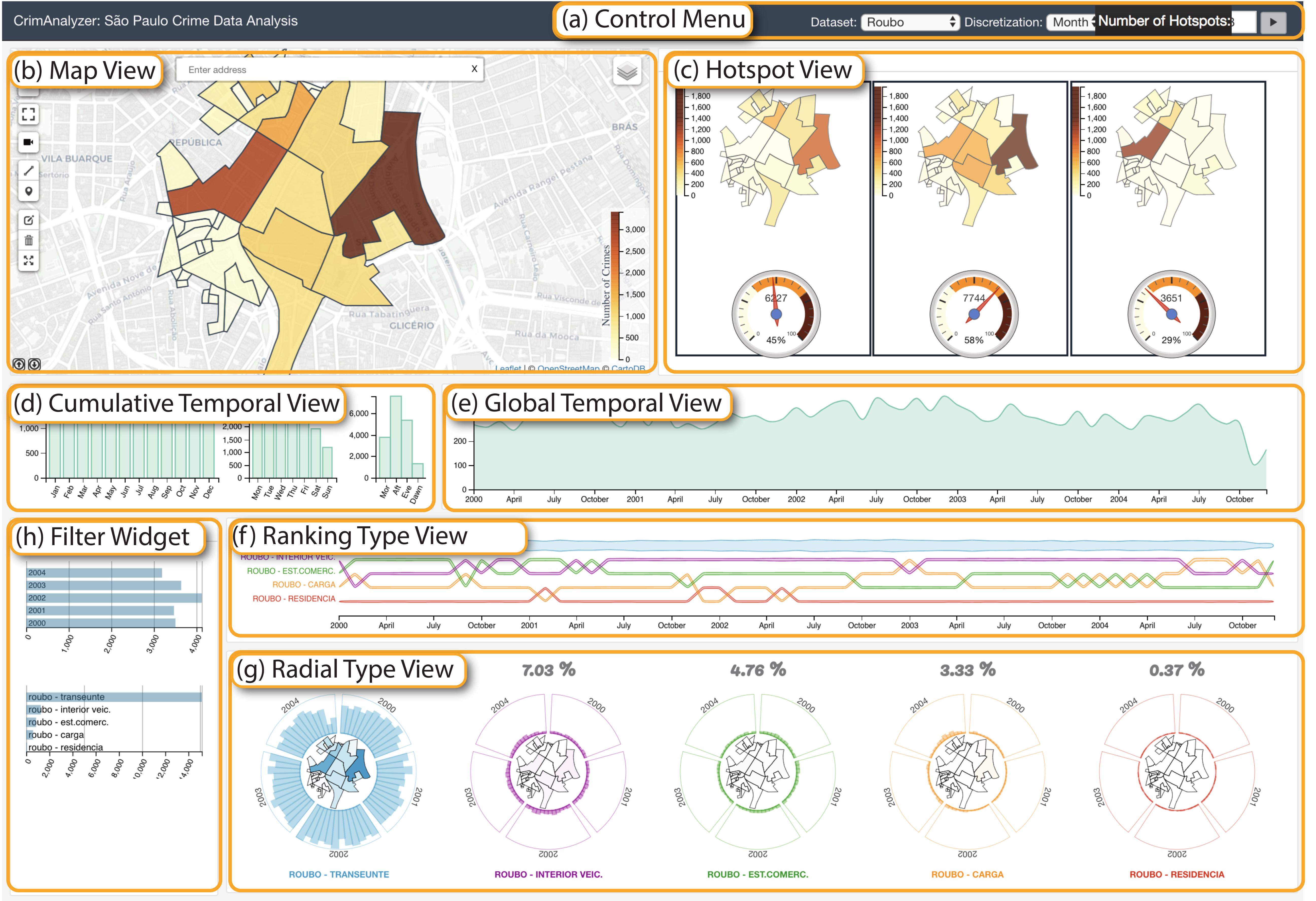}
  \vspace{-11pt}
  \caption{CrimAnalyzer system: the spatial and temporal interactive views enable the exploration of local regions while revealing their criminal patterns over time.\label{fig:teaser}}
}

\vgtcinsertpkg

\begin{document}
\maketitle
\section{Introduction}

Since the mid-1970s, Brazilian society has experienced a known transition process from military dictatorship to democracy.
With this political transition, it was expected that the conflicts would increasingly be solved, reducing the use of violence. 
That was not what happened. The transition was accompanied by a real explosion of conflicts, many of which associated with urban crimes.
There is still no consensus among social scientists about the reasons that explain these trends in the evolution of crime and violence 
in Brazilian society, mainly in the big cities~\cite{adorno2013democracy}. 
Among the explanations that arise more frequently is the exhaustion of traditional security policies models.
Concerning this last aspect,
it is undeniable that crimes have not only grown, but also become more violent and modernized. 
In contrast, agencies in charge of law and order control -- police and criminal justice system -- did not follow these trends and has aged. 
The gap between the dynamics of crime and violence and the state's ability to contain them within the rule of law has widened.
Therefore, introducing modern instruments for the management of public order and crime containment is imperative to make public security policies more efficient, not only in S\~{a}o Paulo but also in any big city in under-development countries. 

In this context, \emph{Crime Mapping}, a branch of Geographic Information Systems (GIS) devoted to explaining 
spatio-temporal behavior of crimes, has emerged as a research field to support criminologists in their analytical process,
leveraging the importance of local geography as a determinant of crime types and the rate they occur in a particular region~\cite{Townsley2017}.
The capability of identifying and visualizing crime hotspots and the ability to filter crime-related attributes to
reveal particular information such as burglary in commercial areas or the seasonality of auto theft in certain neighborhoods are among the key components of a crime mapping approach~\cite{Ratcliffe2010}.
Most existing tools developed in the context of crime mapping focused on the detection of \emph{hotspots}, 
that is, areas with a high number of criminal incidents~\cite{eck2005mapping}.
Although sophisticated mechanisms have been proposed to detect hotspots~\cite{Eftelioglu2016}, 
the search for a high prevalence of crimes ends up neglecting sites where certain types of crimes are frequent but not so intense to
be considered statistically significant~\cite{wang2013understanding}. 
Moreover, most techniques enable only rudimentary mechanisms to analyze an important component of unlawful activities, the temporal evolution of crimes and their patterns. In fact, visualization resources 
for temporal analysis available in the majority of crime mapping systems are very restrictive,
impairing users from performing elaborated queries and data exploration~\cite{andresen2017mapping}.

There is yet another important aspect to be considered in the context of crime mapping, 
the specificities of urban areas under analysis. 
S\~{a}o Paulo city, for example, bears one of the highest criminality rates
in the world, at least one order of magnitude higher than cities such as New York and San Francisco, making glyph-based 
crime mapping solutions such as \emph{LexisNexis}\urlfootnote{communitycrimemap.com}, \emph{NYC Crime Map}\urlfootnote{maps.nyc.gov/crime/}, 
and \emph{CrimeMapping}\urlfootnote{www.crimemapping.com} completely unsuitable
for analyzing crimes in S\~{a}o Paulo.
Nevertheless, the pattern of crimes changes considerably around S\~ao Paulo city, even between regions that are 
geographically close to each other, demanding analytic solutions tailored to scrutinize local regions towards 
revealing their hotspots and crime patterns, while still being able to uncover their dynamics over time. Those
capabilities are not currently available in most crime mapping solutions.

This work presents CrimAnalyzer, a new visualization assisted analytical tool customized to support the analysis of criminal activities in urban areas according to the characteristics of S\~ao Paulo city, 
that is, high criminality rates with great variability in the pattern of crimes, even in geographically close regions. 
CrimAnalyzer enables a number of linked visual tools tailored to reveal patterns of crimes and
their evolution over time, assisting domain experts in their decision-making process and 
providing guidelines not only for repressive but above all preventive actions,
strengthening the planning and implementation of institutional actions, especially from the police. 

In collaboration with a team of domain experts, we have designed visual analytic functionalities 
that allow users to select and scrutinize regions of interest in terms of their hotspots,
crime patterns, and temporal dynamics. Moreover, the proposed system enables resources
for users to dig deeper in particular sites to understand its prevalent crimes and their behavior along time. 
The need for a methodology as implemented in CrimAnalyzer, which 
is able to operate and explore target areas, was one of the main demands raised by the domain experts.
Moreover, CrimAnalyzer implements a methodology based on Non-negative Matrix Factorization~\cite{lee2001algorithms} to 
identify hotspots based not only on the number of crimes but also on the rate they occur.

In summary, the main contributions of this work are:
\squishlist
\item A new methodology to identify crime hotspots based not only on the number of crimes but also 
    on their variation and recurrence rate.
\item A visual analytics machinery that allows users to visually perform spatial and temporal 
    queries towards understanding patterns and temporal dynamics of crimes. 
\item CrimAnalyzer, a visualization-assisted tool that integrates the analytical 
    machinery in a set of linked views. CrimAnalyzer operates on target spatial regions to
    uncover relevant information of the region as a whole and also from its individual sites.
\item A set of case studies revealing interesting phenomena about the dynamics of criminality
    in S\~{a}o Paulo city, supporting hypothesis and theories raised by domain experts and described
    in the literature. 
\squishend

\section{Related Work} 
\label{sec:RW}
The literature about crime analysis is extensive, ranging from statistics and data science to visualization and GIS.
Broadly speaking, crime analysis methods can be grouped into two major categories, geo-referenced and non-geo-referenced approaches. 
The latter, non-geo-referenced approaches, rely on mathematical and computational mechanisms 
such as data mining~\cite{Chen:2004:CDM,Ying:2016}, optimization~\cite{wang2013understanding}, machine learning~\cite{Wang:2013b,Yadav:2017}, statistics~\cite{osgood2010statistical}, and data visualization~\cite{Gao:2014,Xu:2005:CNA,Jentner:2016,Jentner:2018,jackle2017interpretation,xiang2005visualizing}, 
to identify crime patterns, criminal behavior, and also the consistency of criminal justice. 
In the following, though, we focus on geo-referenced techniques developed in the context of crime mapping, 
which are more closely related to the proposed approach.
In order to better contextualize our methodology, we divide geo-referenced techniques in two groups, hotspot centered and 
spatio-temporal criminal pattern identification. It must be clear that there is 
a considerable overlap between those groups, meaning that a hotspot centered technique can also rely
on spatio-temporal patterns to leverage its analysis, but the main focus of such technique is, in fact, hotspot identification.

\myparagraph{Hotspot centered.}
Identifying crime hotspots is a major task in the context of crime mapping~\cite{GIS_Book:2005,eck2005mapping,Eftelioglu2016}.
Although some works rely on Kriging~\cite{oliver1990kriging}, the most common approach for hotspot identification
is a combination of Spatial Scan Statistics~\cite{kulldorff1997spatial} and Kernel Density Estimation (KDE)~\cite{Chainey:2008},
using point clouds, density map, or grid-based approaches as visualization resources~\cite{Nakaya:2010,Johansson:2015,Neto:2016,Santos:2016}. 
As pointed out by Hart and Zandbergen~\cite{Hart:2014}, properly setting the parameters of a KDE is not easy and a  
loose choice of parameters can lead to erroneous or inaccurate results that overestimate or 
disregard hotspot locations~\cite{Chainey:2002}. 
Another issue with KDE based techniques is that locations present regular, but not intense, criminal activities
are hardly pointed out as hotspots. To avoid the issues above, our approach relies on Non-negative Matrix Factorization to
detect hotspots, thus avoiding parameter tuning while being able to ``capture''  locations difficult to be
identified as hotspots by KDE based methods. 

\myparagraph{Spatio-temporal pattern identification.}
Besides crime hotspot identification, the analysis and visualization of temporal and spatial crime patterns are also of great importance
in crime mapping~\cite{Ratcliffe:2006,Brunsdon:2007,Nakaya:2010,Li:2017,Melo:2017}.
Ratcliffe~\cite{Ratcliffe:2004:HM} and Townsley~\cite{Townsley:2008}, for instance, incorporates \emph{aoristic analysis}~\cite{Ratcliffe:2000,Ratcliffe:2002} in their hotspot visualization systems in order to identify important spatio-temporal patterns of crimes. 
Aoristic analysis takes into account the uncertainty of the exact moment that an event occurred when examining the 
overall incidence of crimes over time. 
Lukasczyk et al.~\cite{Lukasczyk:2015} provide a topological perspective of the temporal evolution of hotspots based on 
a spatio-temporal Reeb graph built from a KDE mapping. 
Although interesting, techniques described above are still incipient in clearly revealing spatio-temporal crime patterns and their dynamics.
Our approach, in contrast, combines a number of intuitive visual resources from which one can clearly identify
crime patterns in specific locations and their temporal evolution.

There is a number of spatio-temporal techniques that rely on clustering methods to group 
spatially and/or temporally similar crime events in order to identify patterns.
Those methods can be organized into two categories, the ones that build upon conventional clustering algorithms and 
the ones that rely on Self-Organizing Map (SOM) to identify patterns.
Clustering-based methods extract feature vectors from spatial and temporal crime attributes and 
cluster those attributes via k-means~\cite{Aljrees:2016,Silva:2017} or nearest neighbor clustering~\cite{Kerry:2010,Levine:2013}.

Techniques that rely on SOM have as main goal the identification of similarities among crime attributes.
Chen et al. \cite{Chen:2003:CVC}, in collaboration with the Tucson Police Department, proposed a spatio-temporal 
visualization system called COPLINK, which combines hyperbolic trees, GIS, and SOM in a unified analytical tool.
Andrienko et al.\cite{Andrienko:2010} rely on a SOM matrix display to leverage a visual analytic framework to explore
spatio-temporal similarities between the events.
Hagenauer et al.~\cite{Hagenauer:2011} extended the previous approaches to explore the space-time evolution of the 
patterns in addition to their demographic and socio-economic attributes.
In order to understand patterns between crime types, 
SOM has also been the basis for the spatio-temporal crime analysis system proposed by Guo and Wu~\cite{Gu:2013}, which builds upon 
a visualization infra-structure called VIS-STAMP~\cite{Guo:2006} that integrates dimensionality reduction and parallel coordinates 
in the analysis of crime patterns. 
SOM has well known issues such as the proper setting of weights, number of nodes, and overfitting~\cite{Lampinen:2002}. 
Moreover, SOM-based techniques described above do not integrate hotspot detection as part of the system,
leaving aside an important component of analysis in the context of crime mapping.
%


\section{Challenges, Data Set, and Analytical Tasks} 
\label{sec:challenges_dataset}
In the last eighteen months, we have been interacting with two sociologists experts in social sciences with focus on criminal analysis. 
In particular, one of the sociologists is a researcher with a great reputation in the study of violence in South America. 
The other sociologist is an expert in public safety and social sciences applied to urban issues, 
with background in GIS and large experience in spatio-temporal analysis of crime.
In partnership with the police department of S\~{a}o Paulo state, the team of experts built a data set 
(detailed in Sec.~\ref{sec:data_set}) containing seven years of criminal records in S\~{a}o Paulo city, 
and they invite us to develop a visual analytic tool to assist the understanding and analysis of the data. 

\myparagraph{Nomenclature.} Before further detailing the problem, the requirements raised from the interaction with the domain experts,
and the data set, lets first settle some nomenclature that will be employed throughout the manuscript.

\begin{wrapfigure}{r}{0.19\textwidth}
    \vspace*{-0.33cm}
    \hspace*{-0.25cm} \includegraphics[width=0.19\textwidth]{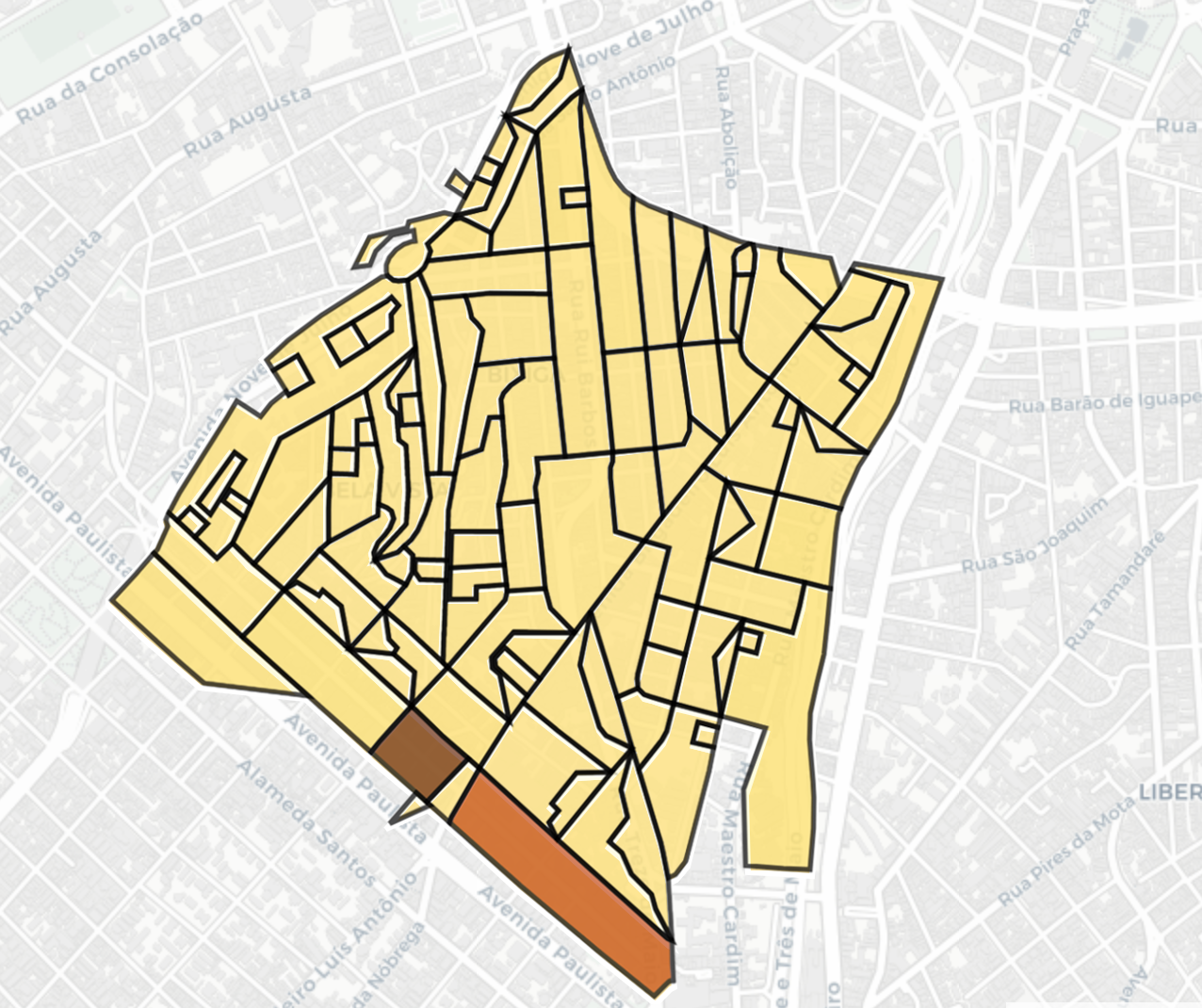}
    \vspace*{-0.4cm}
\end{wrapfigure}
\noindent -- \emph{Site} is the smallest territorial unity given in the spatial discretization. 
        In our context, the sites are defined as the census units of S\~{a}o Paulo city,
        each containing from $250$ to $350$ residences and/or commercial establishments.
        
\noindent -- \emph{Region} is a set of sites, which can correspond to a whole neighborhood, 
        a particular portion of a neighborhood, or even a group of sites adjacent to a street or avenue.
        The inset on the right shows an example of a region and its corresponding sites.

\noindent -- \emph{Hotspots} are sets of sites within a region with relevant criminal activity. 
        The exact meaning of ``relevant'' will be clear when we present the mechanism 
        we designed for hotspot detection. The reddish sites in the inset image correspond to hotspot sites in the given region. 

\noindent -- \emph{Crime type} refers to the type of criminal activity, ranging from burglary to bodily injury
        (death, sexual, and drug-related crimes are not included in our study).  

\noindent -- \emph{Crime pattern} accounts for the prevalence of a group of crime types in a given region or a particular set sites. 
        In other words, if we say that the crime pattern in a set of sites is robbery, car theft, and commercial establishment attack,
        we mean that the three crime types are the most prevalent ones in those sites. 
        
\subsection{Problem Analysis}
\label{sec:pa}
We run several rounds of meetings and interviews with the experts to figure out the main challenges involved in the analysis of crime data. 
After several interactions, we come up with the following issues:
\squishlist
    \item \textbf{Analyzing the characteristics and dynamics of crimes in particular regions of the city.} 
        From their experience and interaction with officers from the police department, the experts conjecture that
        the type and dynamic of crimes have been changing over the years, mainly in particular regions of the city. 
        Moreover, the type of crimes can change even in regions located close to each other depending on the 
        urban characteristics of each region.
        The main difficulty to perform this analysis without a visualization assisted tool is 
        to properly query the data set, which can be a time-consuming and exhaustive job. Many times a large number of images are generated as results, and the work of analyzing them becomes impossible. Moreover, highly prevalent crimes 
        overshadow the presence of less frequent ones, which might also be of interest, demanding specific tools to 
        enable a proper analysis. 
        Face the difficulties, the experts have been performing their analysis
        focused on just one or two types of crime, 
        considering the city as a whole or analyzing large areas that serve as administrative units within the city. 
        Such ``wide'' analysis hampers the validation (or denial) of hypothesis and conjectures that have a local nature. 

    \item \textbf{Identifying crime hotspots within a particular region.}
        The identification of crime hotspots is among the most important tasks when analyzing crimes and their dynamics.
        Hotspots are usually identified as locations that have a greater than the average number of criminal records~\cite{eck2005mapping}.  
        However, criminal sites that are not so prevalent in terms of the number of criminal events, 
        but bears criminal activities that deserve special analysis, tend not to be detected when 
        a ``frequentist'' approach is employed to identify crime hotspots. 
        Due to the lack of more sophisticated mechanisms, the number of criminal records has been the 
        main mechanism employed by the experts in their identification of hotspots. Because of this, it was necessary to propose a new method for hotspot detection that meets the described restrictions. This requirement was emphasized by the domain experts.

    \item \textbf{Understanding and comparing crime patterns.}
        Domain experts believe that sites and hotspots within the same region can present different crime patterns. 
        An issue in this context is to know whether the pattern of crime varies from a site (or hotspot) to another in the same region. 
        In affirmative case, experts would like to understand how crime types are distributed an how they evolve along time.
        The experts were looking for a solution that would intuitively allow them to make such comparisons.
\squishend
Challenges above points to a visual analytic solution endowed with functionalities to easily select regions of interest while
enabling resources to assist the analysis of crime location, patterns, and temporal evolution.
We followed a design process that involved the experts in most stages of the development~\cite{munzner2009nested},
redesigning procedures, components, and functionalities according to experts feedback and demands.

\subsection{S\~{a}o Paulo City Robbery, Burglary, and Larceny Data}
\label{sec:data_set}
The data set assembled by the experts consists of criminal records provided by the police department of S\~{a}o Paulo city.
Only criminal acts as to robbery, burglary, and larceny were provided, leaving out murder, homicide,
drug-related felony, and sexual assault.
\begin{wrapfigure}{r}{0.25\textwidth}
	\vspace*{-0.38cm}
	\hspace*{-0.22cm} \includegraphics[width=0.23\textwidth]{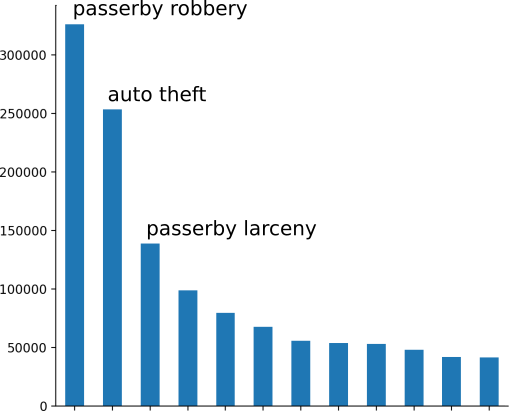}
	\vspace*{-0.35cm}
\end{wrapfigure}
Each record contains the identification number of the census unit (site) where the crime happened, 
the type of crime, and the date and time of the crime.

The data set contains crime records from $2000$ to $2006$. In the very beginning of our 
studies we noticed that the information as to $2005$ and $2006$ was not consistent with previous years and  
a sanity check should be performed by the experts.Since the sanity check turned out more complex than expected, 
we opt to include only information from $2000$ to $2004$ in our studies, in a total of $1,574,920$ records.

Crime types range in $121$ categories, and the $10\%$ most frequent crime types correspond to about $80\%$ of the 
total crimes. The inset on the right shows the histogram of the $10\%$ most prevalent crime types,
labeling the three most frequent ones, \emph{passerby robbery}, \emph{auto theft}, and \emph{passerby larceny}. 
To facilitate the analysis, experts split the original data in three independent categories, 
\emph{vehicle robbery} (includes cars, motorcycles, trucks, etc.) with $295,081$ instances, 
\emph{larceny} in general, with $587,885$ instances,
and a third category with all the other types of robbery and burglary, with $691,954$ instances. 

Although the number of crime types are quite large, the crimes that
domain experts are interested in are not that large, ranging from 3 to 5. Other crime types are sparse enough to be 
analyzed individually, do not demanding a sophisticated visualization tool to interpret them. 
Moreover, some crimes can be grouped into categories, an alternative suggested by the expert and incorporated 
into CrimAnalyzer. In other words,
in each of the three sub-datasets, the experts ranked and grouped the crime types according to their importance.

\subsection{Analytical Tasks}
\label{sec:desiderata}
After identifying the main challenges faced by the experts and understanding how the data
was structured, we conducted a new series of interviews to clearly raise questions to
be investigated. It has become clear that the experts are interested in understanding
the dynamics of crimes over the city by analyzing the variation of crime patterns over
space and time. From the iterative processes with the experts we compiled the following list of analytical
tasks:

\squishlist
\item[\textopenbullet]\textbf{Interactive selections (T1)}:
     How can spatial regions of interest be interactively selected? 
     Is it possible to make the interactive selection of regions flexible enough 
     to pick from single spots to whole neighborhoods and particular avenues?
 \item[\textopenbullet]\textbf{Crime patterns over the city (T2)}:
     Which are the crime patterns in particular regions and sites?. 
     How do criminal patterns change from the center to residential areas and outskirts?
     What about the patterns along the main avenues, streets, and highways?   
\item[\textopenbullet]\textbf{Dynamic of crimes over time (T3)}:
     How have crime types evolved, over time, in particular regions of the city?. More specifically, 
     have crime patterns changed in particular regions over the years?  
     Are crime types seasonal?
 \item[\textopenbullet]\textbf{Crime patterns and hotspots over space (T4)}:
     Which are the hotspots in a given region?
     Which are their crime patterns? How different (if the difference exists) are the crime patterns in distinct hotspots within the same region? 
 \item[\textopenbullet]\textbf{Crime patterns and hotspots over time (T5)}:
     Have crime hotspots changed over time in a given region?
     Have crime patterns changed over time in a given hotspot?
  
\squishend

As mentioned before, the lack of interactive mechanisms to select regions of interest
combined with general-purpose analysis and visualization techniques have prevented domain
experts from freely explore the data to verify hypotheses and conjectures. The
first step to enable more powerful analytic resources is the design of a proper
interactive selection tool, being this the goal of T1.

It has also become clear during the interviews that it is important to drill down from 
high-level summaries to individual analysis of sites and hotspots. Analyzing data in
different scales is also essential to understand how patterns vary across space and time.
For example, the pattern of crimes and hotspots can change throughout months and over
different days of the week. This fact is related to T3, and demands particular data
aggregation and filtering to be properly addressed. 

Analytical tasks T2 and T3 are related to the problem of understanding the different patterns of 
crimes around the city and their evolution over time, as discussed in Sec.~\ref{sec:pa}, while 
tasks T4 and T5 are associated to the problem of analyzing hotspots, also discussed in Sec.~\ref{sec:pa}.
To be properly tackled, those tasks demand specific mechanisms to detect hotspots and also visual resources to explore and understand them.
	
Among our goals is the integration of interactive selection methods and
dedicated visual analysis tools towards allowing domain experts 
to accomplish both confirmatory and exploratory analysis. 
Moreover, some domain experts are not trained in computer science, thus the system 
should be as simple and intuitive as possible. 
However, simplicity and expressiveness must be balanced to render the system capable of supporting
spatio-temporal analysis at different scales, while being able to uncover non-trivial hotspots and
crime patterns.

\subsection{The CrimAnalyzer System} 
\begin{figure}[t!]
  \centering
  \includegraphics[width=.98\linewidth]{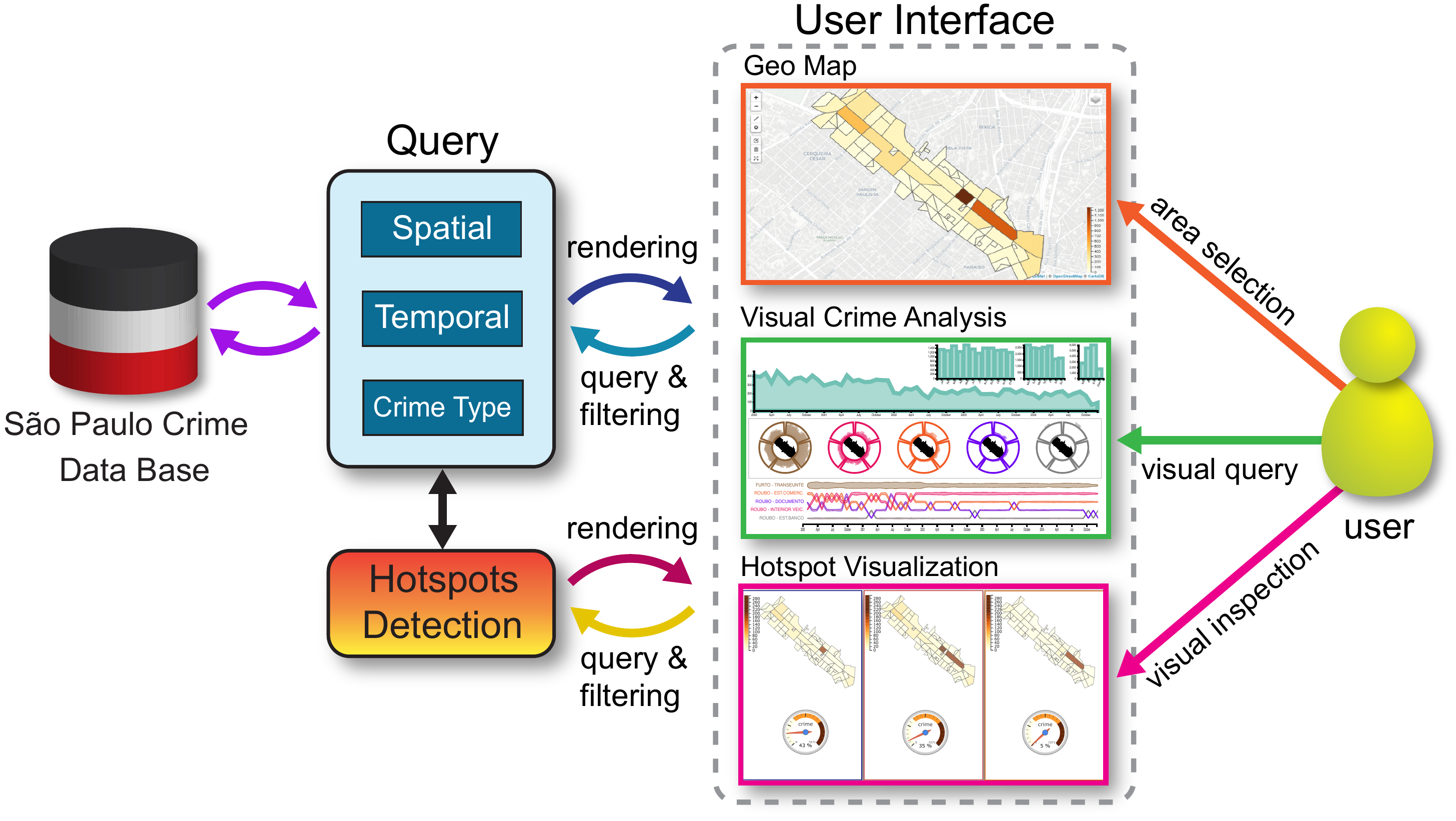}
  \vspace*{-0.2cm}
  \caption{Pipeline overview of the CrimAnalyzer System.}
	\label{fig:pipeline}
  \vspace*{-0.5cm}
\end{figure}

Based on the raised requirements and analytical tasks,
we have developed CrimAnalyzer, a system for exploring spatio-temporal crime data in particular locations.
CrimAnalyzer enables simple, yet compelling, visual resources to query, filter and visualize crime data.
The visual resources are supported by a mathematical and computational machinery tailored to 
extract and polish information so as to visually present it in an intuitive and meaningful way.
The modules and system architecture are illustrated in \autoref{fig:pipeline}.
Users visually query the data set by interacting with a map and selecting a region of interest as well as
by interacting with the different linked views that make up the system.

\section{Hotspot Identification Model}

\begin{figure*}[t!]
	\centering
	\subfigure[Region of Interest]{\label{fig:region_synthetic}\includegraphics[width=0.2\linewidth]{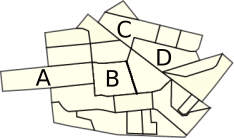}}
	\subfigure[Data matrix $X$]{\label{fig:X}\includegraphics[width=0.22\linewidth]{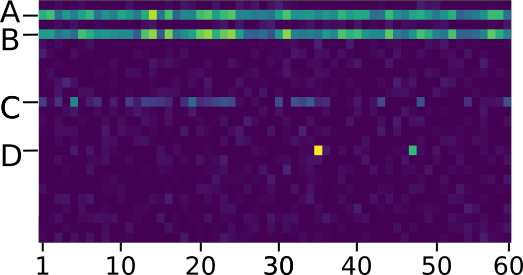}}
	\subfigure[Hotspots (columns of $W$) and their occurrence (rows of H)]{\label{fig:synthetic_WH}\includegraphics[width=0.25\linewidth]{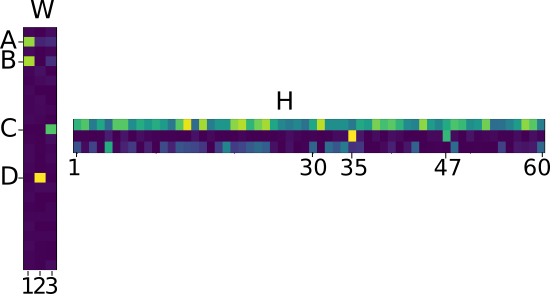}}
	\subfigure[The same as c) but with rank 5]{\label{fig:synthetic_WH_rank5}\includegraphics[width=0.25\linewidth]{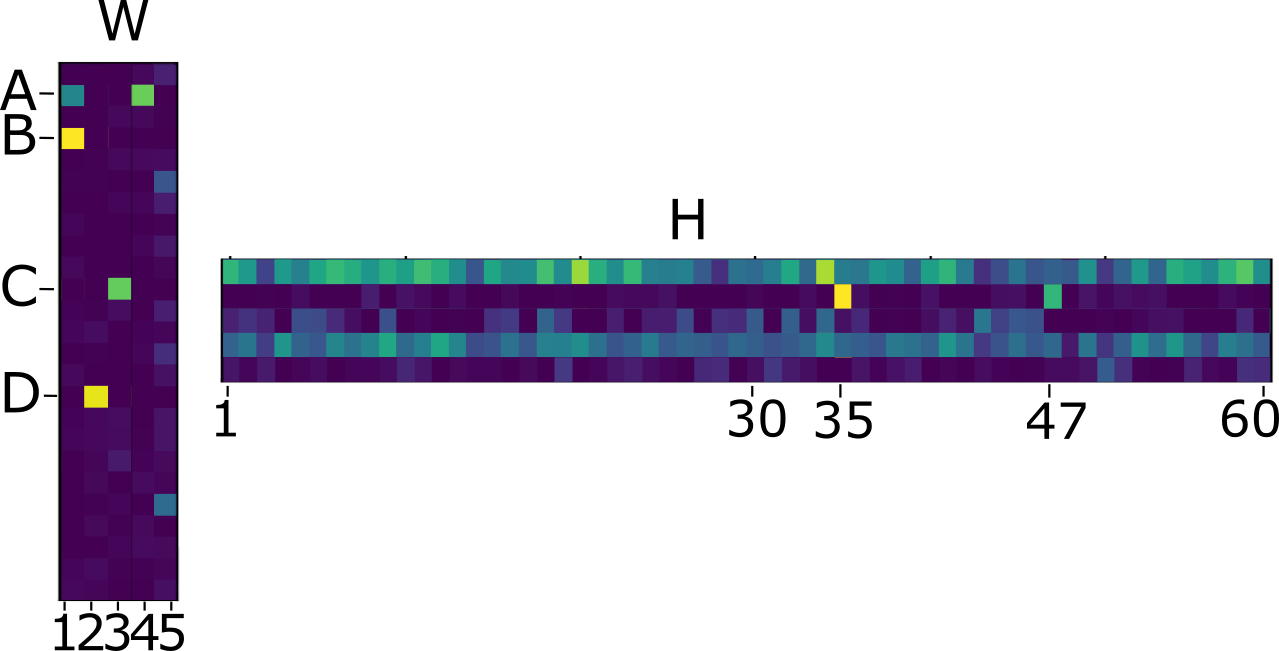}}
	\vspace{-0.2cm}
	\caption{a) Region of interest. b) Data matrix containing crime information from the regions in a). Rows correspond to sites
		while columns are time slices. The darker the color, the closer to zero the number of crimes is. c) Rank 3 NMF decomposition from $X$.
		d) Rank 5 NMF from $X$.}
	\label{fig:synthetic}
\end{figure*}

As discussed in Sec~\ref{sec:challenges_dataset}, hotspot identification is one of the most important tasks 
for crime analysis. 
In our context, hotspots have a more general connotation than in previous works, 
corresponding to sites where criminal activity is high but also to locations where the number
of crimes is not large, but frequent enough to deserve a detailed analysis. For example, 
a region can contain sites where the number of crimes is much larger than in any other sites, which clearly correspond to important hotspots. However, the region can also contain
a particular site where crimes are frequent, but happening in much smaller magnitude 
if compared against the prominent ones.
Moreover, the region can also contain sites where crimes are not frequent at all, but present
spikes in particular time frames.
In our context, we consider the three different phenomena as hotspots, seeking to
identify sites where crimes are frequent and in large number, sites where crimes are
frequent but do not in large number, and sites where crimes are not frequent, but 
happen in large numbers in particular time frames. 
The different crime behavior will be further discussed and illustrated in Sec.~\ref{sec:det_hot}.

\myparagraph{Analysis of individual sites.}
There are many alternatives to identify hotspots, as discussed in Sec.~\ref{sec:RW}.
Although varying in terms of complexity, existing techniques typically rely on the comparison of  
statistical information to identify hotspots. 
Hotspots can be identified as particular points or as area units, depending on how the data is organized,
delegating to the visualization the task of properly reveling the hotspots.
The problem with this approach is that 
crimes happening in small magnitude or in isolated time frames tend not to be statistically significant,
hardly being pointed out as hotspots. 

Another issue  is that several sites might be identified as  
hotspots, but their temporal relation remains unrelated. For example, sites can be timely correlated, meaning that crimes are committed in those sites in the same time slices. It makes sense 
to group timely correlated sites in a single hotspot, but computing hotspots individually and group them 
according to temporal matches is not easy and involves the use of thresholds to decide which sites should be grouped.

\myparagraph{Analysis of groups of sites.}
Instead of analyzing sites individually, one can resort to techniques that directly identifies groups of sites as hotspots. 
A straightforward alternative is to 
extract features from the sites and apply a clustering scheme to group similar sites
in hotspots (see Sec.~\ref{sec:RW}). However, the problem of extracting meaningful features that characterize sites spatially and temporally
is quite involved, mainly due to the sparsity of the crime data.
In fact, in the course of our development, we tried different alternatives to define spatio-temporal crime feature vectors, 
ranging from simple cumulative time windows to more sophisticated methods based on graph wavelet 
coefficients~\cite{valdivia2015wavelet}, but we could not obtain results that complied with our requirements. 

To get around the difficulties pointed above, we opted to an approach based on Non-Negative Matrix Factorization~\cite{lee2001algorithms},
which worked pretty well in our context, identifying hotspots according to our requirements.

\subsection{Non-Negative Matrix Factorization}
Before presenting the details on how we have adapted NMF to operate in our context, lets shortly review the main concepts 
and ideas involved in an NMF analysis. 
An $m\times n$ matrix $X$ is said \emph{non-negative} if all entries in $X$ are greater or equal to zero ($X\geq 0$).
The goal of NMF is to decompose $X$ as a product $W\cdot H$, where $W$ and $H$ are non-negative matrices with dimensions $m\times k$ and 
$k\times n$, respectively (the roles of $m,n$, and $k$ will be clear in Sec.~\ref{sec:det_hot}). 
In mathematical terms, the problem can be stated as follows: 
\begin{equation}
    \arg\min_{W,H} \|X - WH\|^2 \quad \mbox{subject to}\,\,\,\,\, W,H\geq 0
    \label{eq:mnf}
\end{equation}

A solution for the minimization problem (\autoref{eq:mnf}) provides a set of basis vector $w_i$, corresponding
to the columns of $W$, and a set of coefficients $h_j$, corresponding to the columns of $H$, such that each column $x_j$ of $X$
is written as the linear combination $x_j=\sum_i h_{ij}w_i$, (or $x_j=Wh_j$). 
In other words, for each column in $X$ we have a corresponding column in $H$ whose entries are coefficients associated to the columns (basis vectors) of $W$. 
The matrix representation below (~\autoref{eq:nmf_color}) illustrates the relation between columns of $X$ and $H$ as well as
coefficients in $H$ and basis vector in $W$.

\vspace{-0.5cm}
\begin{equation}
\arraycolsep=1.4pt
\left[\begin{array}{cc>{\columncolor{gray!20}}ccc}
    | & & \vert & & | \\
    x_1 & \dots & x_j & \cdots & x_n \\
    | & & \vert & & | 
\end{array}\right]
=
    \left[\begin{array}{>{\columncolor{orange!30}}c>{\columncolor{yellow!30}}cc>{\columncolor{blue!30}}c}
    | & | & & | \\
    w_1 & w_2 & \cdots & x_k \\
    | & | & & | 
\end{array}\right]
\left[\begin{array}{cc>{\columncolor{gray!20}}ccc}
    \multirow{4}{*}{$\begin{array}{c}\rule{.4pt}{4mm} \\  h_1 \\ \rule[-2mm]{.4pt}{4mm} \end{array}$} & \multirow{4}{*}{$\cdots$} & \cellcolor{orange!30} h_{1j} & \multirow{4}{*}{$\cdots$} & \multirow{4}{*}{$\begin{array}{c}\rule{.4pt}{4mm} \\ h_n \\ \rule[-2mm]{.4pt}{4mm} \end{array}$} \\
     & & \cellcolor{yellow!30} h_{2j}  & &  \\
     &  & \vdots &  & \\
     & & \cellcolor{blue!30} h_{kj} & & 
\end{array}\right]
    \label{eq:nmf_color}
\end{equation}

There are two important aspects in an NMF decomposition that will be largely exploited in the context of
hotspot detection, namely, low rank approximation and sparsity.
Low rank approximation accounts for the fact that the basis matrix $W$ usually has much lower 
rank than the original matrix $X$, meaning that (the columns of) $X$ is represented using
just a few basis vectors. 
As detailed in the next subsection, we rely on low rank approximation to define the number of hotspots, that is, 
by setting the rank of $W$ we also set the number of hotspots. 
Sparsity means the basis and coefficient matrices contain many entries equal (or close) to zero, 
which naturally enforces only relevant information from $X$ to be kept in $W$ and $H$. 
This fact is important to identify particular sites within a hotspot and the time slices
where each hotspot shows up.

\subsection{Identifying Hotspots with NMF}
\label{sec:det_hot}
We rely on NMF to identify hotspots, their rate of occurrence and ``intensity''. The matrix $X$ to be decomposed as the product $W\cdot H$
comprises crime information in a particular region of interest. 
Specifically, each row in $X$ corresponds to a site of the region and each column to a time slice. 
In~order to facilitate discussion, we present the proposed approach using a synthetic example.
\autoref{fig:region_synthetic} shows a region made up of $25$ sites and we generated 
synthetic crime data in $60$ time slices, representing months over five years.
For sites denoted as \textsf{A} and \textsf{B}, we draw $60$ samples
from a normal distribution with mean $8$ and variance $4$,
ensuring that \textsf{A} and \textsf{B} are correlated, that is, when the number of crimes in \textsf{A} is large the same happens with \textsf{B} 
(the number of crimes in \textsf{B} is generated by perturbing the values of \textsf{A} using a uniform random distribution
with values between $-3$ and $3$).
This construction is simulating two regions with high prevalence of crimes over time. 
Crimes in the site denoted as \textsf{C} in \autoref{fig:region_synthetic} follows a normal distribution
with mean $1$ and variance $4$, corresponding to a location where crimes are not large in number, but happening quite frequently.
Finally, for site \textsf{D} we draw $60$ samples from a normal distribution with mean $0$ and variance $0.25$, except
for time slices $35$ and $47$, where we set the number of crimes equal to $15$ and $10$ respectively, simulating
a site where crimes are no frequent, but happen in large numbers in particular time slices.
For all the other sites we associated $60$ samples drawn from a normal distribution with mean $0$ and variance $0.25$. 
Values for all sites are rounded to the closest integer and negative values set to zero.
\autoref{fig:X} illustrates the matrix $X$ built from the synthetic data described above.
Notice that the simulated crime dynamics is clearly seen in $X$.

Given an $m \times n$ matrix $X \geq 0$, an NMF decomposition of $X$ results in matrices $W \geq 0$ and $H \geq 0$. 
In practice, the rank of $W$  is significantly less than both $m$ and $n$, i.e., $k = \rank(W) \ll m,n$. In~our context, the columns of $W$ 
correspond to hotspots while entries in the rows of $H$ indicate the ``intensity'' of the hotspot in each time slice.
\autoref{fig:synthetic_WH} illustrates matrices $W$ and $H$ obtained from matrix~$X$ in \autoref{fig:X} using
a NMF decomposition with rank $k=3$. Notice that the entries in first (left most) column of $W$ have values close to 
zero almost everywhere, except in the entries corresponding to the sites \textsf{A} and \textsf{B}. Therefore, the hotspot  
derived from the first column of $W$ highlights sites \textsf{A} and \textsf{B} as the relevant ones. 
The high prevalence of crimes on those
regions can clearly be seen from the first (top) row of matrix $H$, which has most of its entries with non-zero values. 
The second column of $W$ is mostly null, except in the entry corresponding to site \textsf{D}, where crimes are not
frequent but happen with high intensity in particular time slices. Notice that the second row of $H$ has basically
two entries different from zero, corresponding exactly to the time slices $35$ and~$47$, when the site \textsf{D} faces a large number
of crimes. Finally, the last column of $W$ gives rise to a hotspot that highlights site \textsf{C}, where crimes are frequent,
but in smaller magnitude when compared to \textsf{A} and~\textsf{B}. The~incidence and intensity of crimes in \textsf{C} are clearly seen in the third (bottom)
row of $H$.

One can argue that the results presented in~\autoref{fig:synthetic_WH} worked so well because we wisely set the 
rank of $W$ equal $k=3$ and that in practice it is difficult to find a proper value for the rank.  
To answer this question, \autoref{fig:synthetic_WH_rank5} shows the result of factorizing matrix $X$
setting the rank of $W$ equal $k=5$. Notice that the main difference between the rank $k=3$ and rank $k=5$ 
factorizations is that the first column of $W$ in \autoref{fig:synthetic_WH} was split into two columns 
in the rank $k=5$ factorization, giving rise to columns $1$ and $3$ of $W$ in \autoref{fig:synthetic_WH_rank5}.
Nevertheless, the third column still indicates the correlation between \textsf{A} and \textsf{B}, which thus is 
not completely missed due to the presence of column $1$. The right most column of $W$ in \autoref{fig:synthetic_WH_rank5}
is mostly noise and it represents sites with a little criminal activity, what is attested
by the bottom row of $H$ in \autoref{fig:synthetic_WH_rank5}, which is almost null. 
Therefore, increasing $k$ tends to split meaningful hotspots while creating some 
noisy, not so important ones, which can easily be identified from almost zero rows in $H$.

\myparagraph{Improving the identification of hotspots.}
Most entries in matrix $H$ are close to but are not zero, demanding a threshold 
 to decide whether or not a hotspot takes place in a given time slice. Playing with thresholds
is always inconvenient, mainly for non-experienced users. In order to avoid the use of thresholds, we binarize the matrix $H$ using the Otsu's algorithm~\cite{otsu1979threshold}, considering that a hotspot appears in
a given time slice if its corresponding entry in the binarized version of matrix~$H$ is $1$. 

The synthetic example discussed above shows that hotspots generated from NMF attend the
requirements of our problem, justifying our choice of NMF as mathematical model for tackling
the problem. Among the different versions of NMF, we opt to the sparse
non-negative matrix factorization proposed by Kim and Park~\cite{kim2007sparse}, which allows for
enforcing sparsity in both $W$ and $H$ simultaneously.

We conclude this section saying that, as far as we know, this is the first work to employ
NMF as a mechanism to detect hotspots in the context of Crime Mapping.

\myparagraph{Comparison with spatial statistics}
The \emph{Getis-Ord $G_i^{*}$ statistics}~\cite{Gi:1992,ord1995local} is a well-known hotspot detection method 
available in the toolbox \emph{Local Indicator of Spatial Association}~(LISA) ~\cite{Anselin:1995}
$G_i^{*}$ operates by measuring the local spatial autocorrelation variation over a region of interest.
$G_i^{*}$~reports a p-value and a z-score for each location in the region of interest, 
marking as hotspots those with statistically significant (low p-values) large z-scores.

\begin{figure}[!ht]
	\centering
	\subfigure[S\~ao Paulo city clustering]{\label{fig:sp_division}\includegraphics[width=0.54\columnwidth]{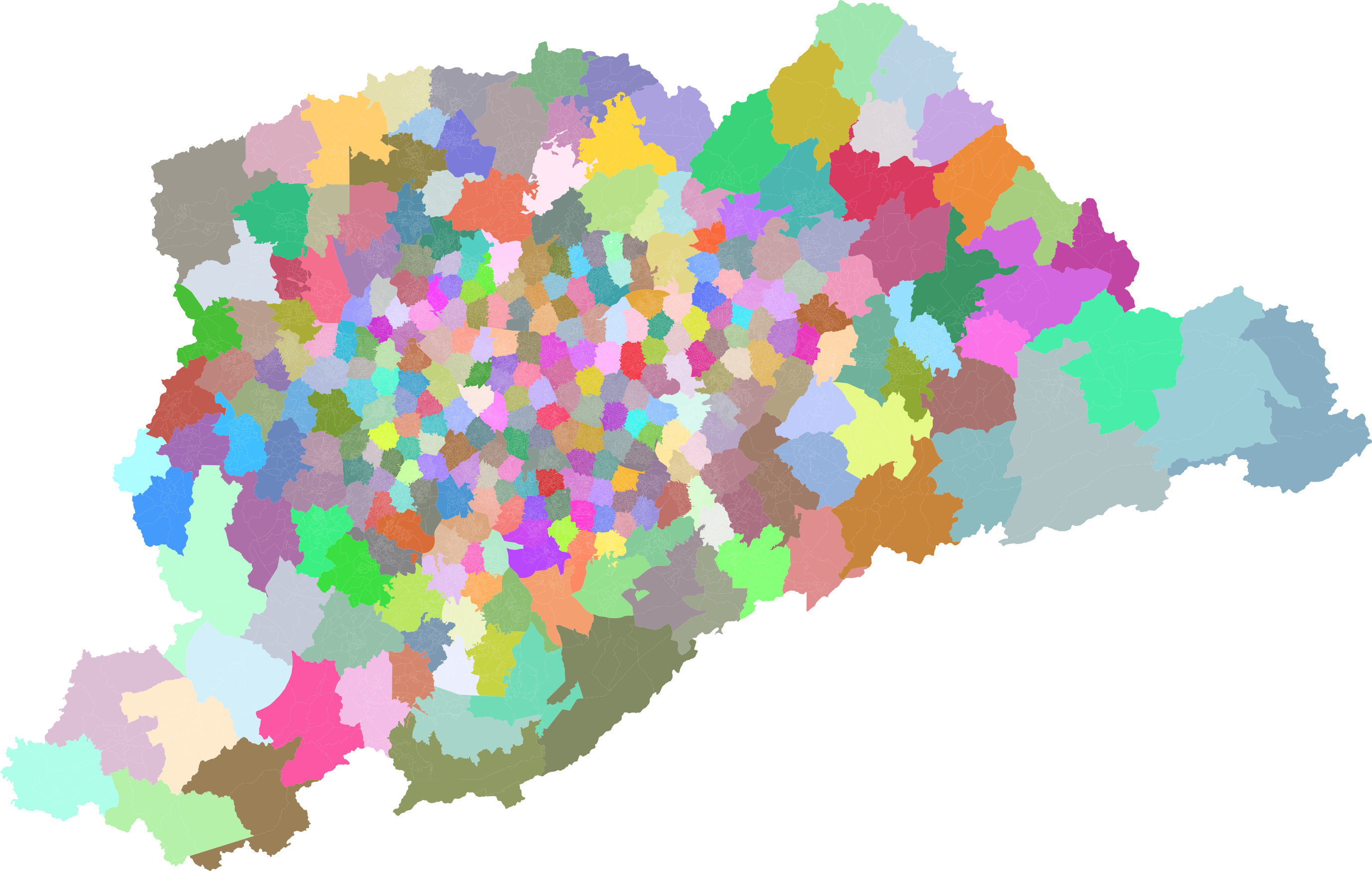}}
	\subfigure[$SSI$ distribution for $k=3$]{\label{fig:k3_SSI}\includegraphics[width=0.45\columnwidth]{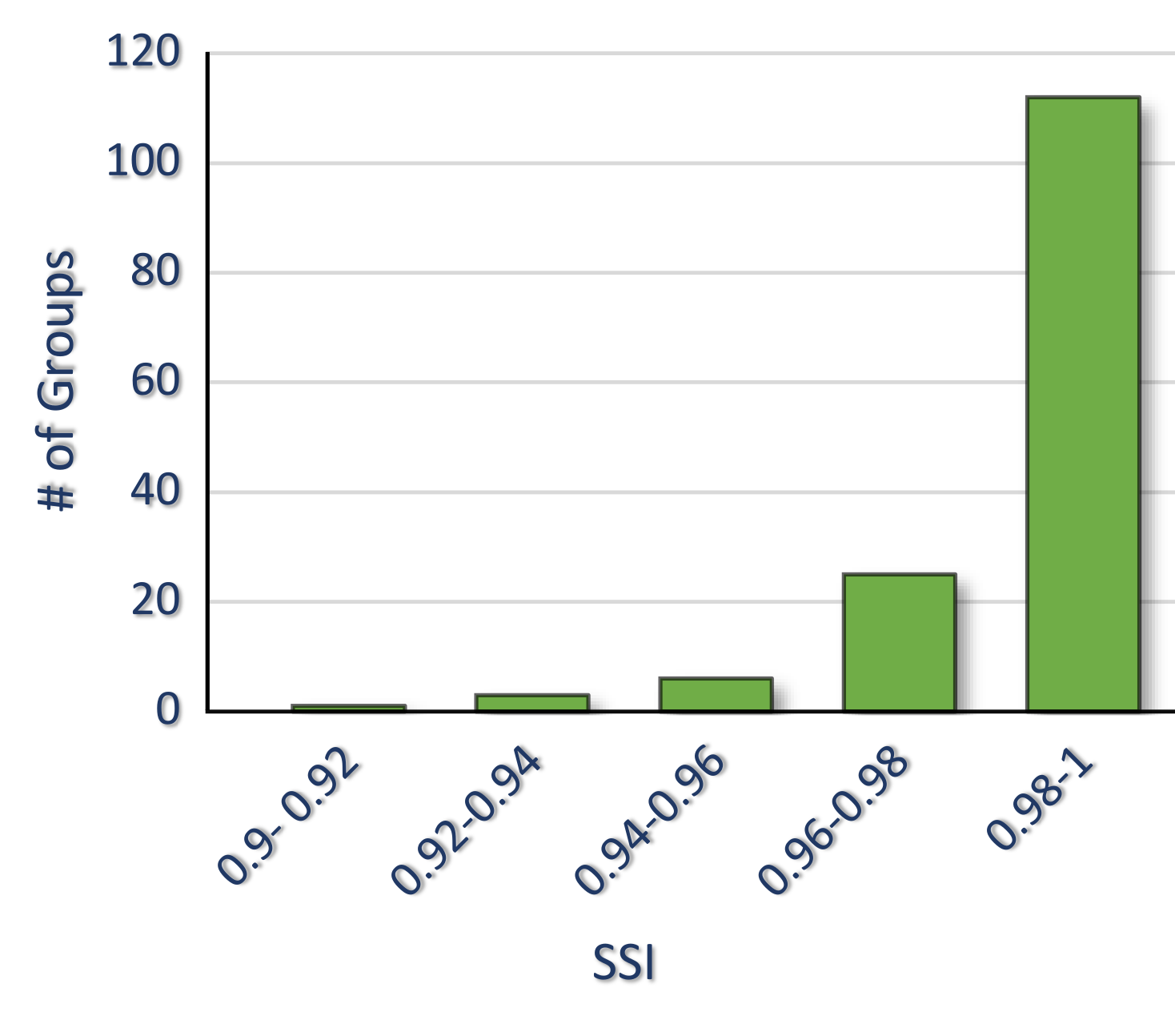}}
	\caption{(a) Division of S\~ao Paulo city into $300$ groups and (b) SSI distribution in those regions. NMF and $G_i^{*}$ detect the same hotspots in most of the cases.}
	\label{fig:SaoPauloResult}
\end{figure}

In order to perform a quantitative comparison between NMF and $G_i^{*}$, we grouped the census units into $300$ 
regions as shown in Fig.~\ref{fig:sp_division}. 
{The regions have been computed by applying K-means to the coordinates of the centroid of the census units.  
Since there are about $30,815$ census units, setting the number of clusters equal to $300$ tends to generate groups with about $100$ units in
the denser areas of the city.}
{The \emph{Sokal-Sneath index}~(SSI)}, a well known binary data classification similarity measure~\cite{Choi10asurvey},
is employed to compare the hotspots resulting from NMF with $k=3$ (the default rank value in our system) 
against the ones obtained by~$G_i^{*}$ with a 99\% confidence level 
(we relied on the $G_i^{*}$ implementation available in the \emph{PySAL}~Python library~\cite{pysal2007}).
Specifically, we~assign each site to one of the four categories (labels):

\squishlist
	\item $\mathsf{P}$: if the site is a hotspot for both NMF and $G_i^{*}$ (positive match);
	\item $\mathsf{F}$: if the site is a hotspot detected by NMF, but not by $G_i^{*}$;
	\item $\mathsf{G}$: if the site is a hotspot detected by $G_i^{*}$, but not by NMF;
	\item $\mathsf{N}$: if the site is not a hotspot for both methods (negative match).
\squishend
The SSI similarity measure is then computed as:
$$
SSI = \frac{2|\mathsf{P}| + 2|\mathsf{N}|}{2|\mathsf{P}| + |\mathsf{F}| + |\mathsf{G}| + 2|\mathsf{N}|} \,,
$$
where $|\cdot|$ denotes the cardinality. A $SSI = 1$ means that the hotspots detected by both methods in a given region match exactly.

Histogram depicted in Fig.~\ref{fig:k3_SSI} gathers SSI values from all the $300$ regions.
Note that $SSI$ values are larger than $0.90$, most of them lying in the range $[0.98,1.00]$, showing 
the good match between NMF and $G_i^{*}$.
In fact, most of the locations pointed out as hotspot by $G_i^{*}$ are also captured by NMF.
However, in about $200$ regions, NMF detected a few more hotspots than $G_i^{*}$.

\begin{figure}[!t]
	\centering
	\subfigure[$|\mathsf{F}| = 2$ and $|\mathsf{G}| = 0$] {\label{fig:nmf_mas_lisa}\includegraphics[width=0.39\columnwidth]{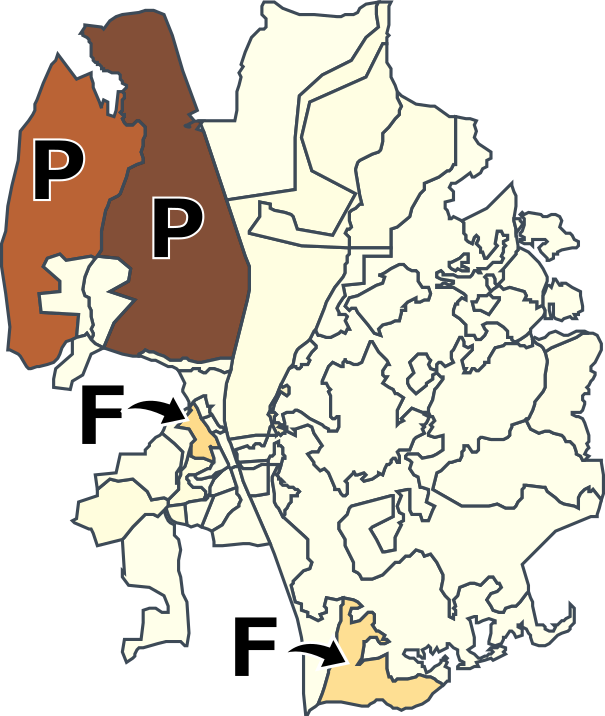}}\hspace{1.5cm}
	\subfigure[$|\mathsf{F}| = 0$ and $|\mathsf{G}| = 2$] {\label{fig:lisa_mas_nmf}\includegraphics[width=0.4\columnwidth]{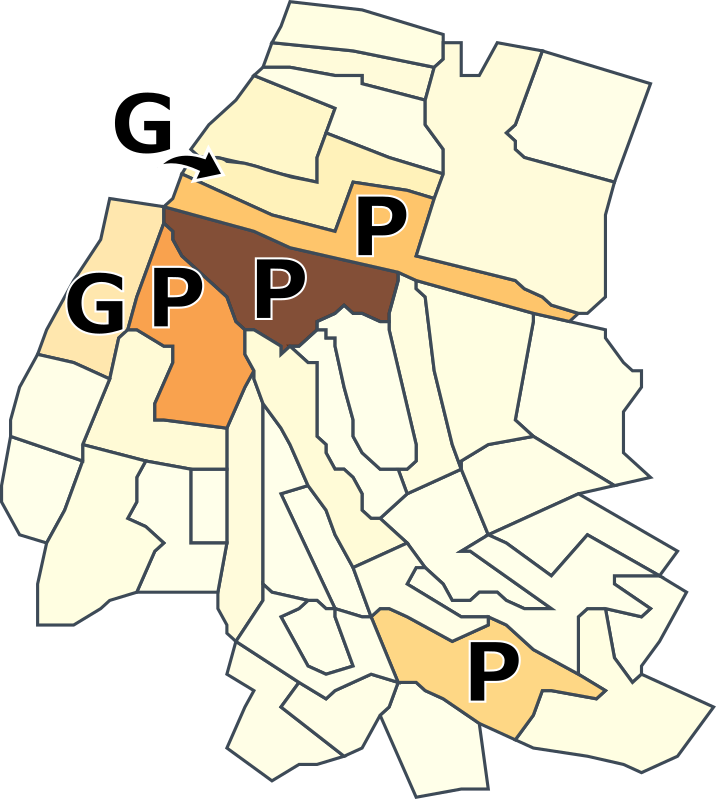}}
	\caption{
	Qualitative comparison between NMF and  $G_i^{*}$. (a) Region where NMF detects more hotspots and (b) Region where $G_i^{*}$ detects more hotspots.
	}
	\label{fig:hotspotdiscussion}

\end{figure}

Fig.~\ref{fig:hotspotdiscussion} illustrates typical situations where NMF and $G_i^{*}$ differ in a few unities. 
In Fig.~\ref{fig:nmf_mas_lisa}, NMF and $G_i^{*}$ have both found the hotspots labeled as $\mathsf{P}$ 
(the labels in Fig.~\ref{fig:hotspotdiscussion} are according to the classification used by SSI, the darker the site is, the more dangerous it is), 
but NMF has captured two extra hotspots, labeled as $\mathsf{F}$ (unlabeled units belong to the category $\mathsf{N}$).
Notice that the color code indicates that the sites $\mathsf{F}$ are indeed regions with a prevalence of criminality,
although not captured by $G_i^{*}$.
In Fig.~\ref{fig:lisa_mas_nmf}, in contrast, $G_i^{*}$ detects two more hotspots than NMF ($\mathsf{G}$~sites).
Notice that the color code of the units marked as $\mathsf{G}$ in Fig.~\ref{fig:lisa_mas_nmf} indicates that 
crimes in those regions are not so intense as in the $\mathsf{P}$ hotspots. The reason why $G_i^{*}$
points the $\mathsf{G}$ sites as hotspots is that those sites are neighbors of units where the number of crimes is high
(``real'' hotspots), so the kernel integration employed by $G_i^{*}$ ends up being contaminated
by the neighbor sites where crimes are prevalent. In other words, the $\mathsf{G}$ sites
are pointed out as hotspots due to their proximity with $\mathsf{P}$~sites. 
Sites pointed as $\mathsf{F}$ in  Fig.~\ref{fig:nmf_mas_lisa} have not been captured by $G_i^{*}$ 
because they are isolated in the middle of units with no crimes. Therefore, besides not demanding a grid discretization,
NMF tends to capture hotspots in a more consistent manner, being an attractive alternative to conventional statistical approaches.

The value $k$ (NMF rank) impacts the $SSI$ measure. We have run the comparisons ranging $k=3,\ldots,10$, getting
an average $SSI$ greater than $0.98$ for $k=3,4,5$, but slightly better for $k=3$. This result motivated us to set $k=3$ as the
default value in CrimAnalyzer.

\section{Visual Design}
This section describes the visual components of CrimAnalyzer. \autoref{fig:teaser} illustrates 
the web-based system, which comprises a Control Menu (a) six interactive views (b-g), and a filter widget (h).
\autoref{tab:views} shows the properties of each view. For instance, \emph{Map View} shows the space facet and 
\emph{Ranking Type View} the temporal and crime type facets.
Columns under \emph{filter}'s category show how to interact with each view. Some views allow to 
constraint space, time and crime types.
	
The design of visual resources was driven in order to satisfy the analytical tasks described in  Sec.~\ref{sec:desiderata}. In \autoref{tab:views}, Columns under \emph{task's} category indicate the relation of each view and the analytical tasks described. For instance, \emph{Ranking Type View} and \emph{Radial Type View} satisfy all analytical tasks.

Some visual resources, as the \emph{Ranking Type View}, has never been used in the context of crime analysis.
As will become clear in Sec.~\ref{sec:caseStudies}, the proposed visual encoding turns out to be quite 
effective to elucidate the dynamics of different types of crime over time in particular locations of the city.

\begin{table}[!t]
  \resizebox{\columnwidth}{!}{%
  \begin{tabular}{lccccccccccc}
    \toprule
  & 
  \multicolumn{3}{c}{Facet} & 
  \multicolumn{3}{c}{Filter} & 
  \multicolumn{5}{c}{Task} \\
  \cmidrule(lr){2-4}
  \cmidrule(lr){5-7}
  \cmidrule(lr){8-12}
  & Space & Time & Type & Space & Time & Type & T1 & T2 & T3 & T4 & T5 \\
  \midrule
  \rowcolor[gray]{.9}
  Map View                 & \ck &     &     & \ck &     &     & \ck & \ck &     & \ck &    \\
  Hotspot View             & \ck & \ck &     &     & \ck &     & \ck &     &     & \ck & \ck \\
  \rowcolor[gray]{.9}
  Cumulative Temporal View &     & \ck &     &     & \ck &     & \ck &     & \ck &     & \ck \\
  Global Temporal View     &     & \ck &     &     & \ck &     & \ck &     & \ck &     & \ck \\
  \rowcolor[gray]{.9}
  Ranking Type View        &     & \ck & \ck &     &     &     & \ck & \ck & \ck & \ck & \ck \\
  Radial Type View          &     & \ck & \ck &     &     & \ck & \ck & \ck & \ck & \ck & \ck \\
  \bottomrule
  \end{tabular}
  }
    \caption{View properties and their analytical tasks (Sec.~\ref{sec:desiderata}).}
  \label{tab:views}
  \vspace{-10pt}
\end{table}

In the following, we describe each visual components, starting with the \emph{Control Menu} (see \autoref{fig:teaser}). 

\subsection{Control Menu}
The \emph{control menu} has three options: dataset, time discretization (\ie, months or days) and the number of hotspots (rank of the NMF decomposition). 
As shown in \autoref{fig:teaser}(a) we are using the dataset ``Roubo'' with monthly discretization and three hotspots (default) 
in most of our analysis. 
The NMF decomposition is performed by the Python library called \emph{Nimfa}~\cite{Zitnik2012}, 
which enables resources to evaluate and automatically choose the rank value $k$.
However, such an automated process is computationally costly, impairing its use in an interactive visual analytic application such as CrimAnalyzer. Therefore, we have opted to allow the user manually set $k$.

\subsection{Map View}
This is the first component used by users to start the analytical process, enabling resources to define the region of interest. This
view is comprised of a geographical map and a choropleth map to encode the number of crimes
committed at each \emph{site} in the region. Also, users can perform interaction to zoom and pan the map. 

\myparagraph{Region selection:} 
Users can define a region of interest by 1) clicking on the map (to select a site), 2) drawing a polyline (to select avenues or streets for example),
3) drawing a polygon (to select a whole neighborhood), or 4) provide the address of a location. 
Drawings can be expanded to comprise other sites in the neighborhood. 
Finally, CrimAnalyzer defines the \emph{region} by computing the \emph{sites} that intersect the drawing. 
In \autoref{fig:teaser}(b) we can see how a region is represented.

\myparagraph{Site selection:} 
During the exploration, when a region has already been defined, this view might be used for spatial filtering (\eg, to focus in a particular site). This operation is performed by clicking a site, which is highlighted by mapping a texture to the corresponding area.

\myparagraph{Filtering:} 
When other views make an spatial filtering (\ie, selecting a site), the corresponding site is highlighted by changing its texture.
When a time or type filter is activated by other views, our choropleth map is recalculated using the filtered data. 

\subsection{Hotspots View}
An important component of our approach is the hotspots identification. 
In Sec.~\ref{sec:det_hot}, we explained how Non-Negative Matrix Factorization has been used to reveal hotspots. In this view, 
we use multiple maps to represent the spatial distribution of each hotspot. Users can specify the number of hotspots in the \emph{Control Menu}. 
Below each hotspot (see \autoref{fig:teaser}(c)), there is a gauge widget that depicts the number of crimes in the hotspot 
(the top number in the gauge),
the temporal rate of occurrence of the hotspot (the bottom percentage in the gauge), and how relevant is that hotspot in the whole set of crimes
(the gauge pointer).
The importance of the hotspot is computed by a function $f:[0,1]\times[0,1]\rightarrow [0,1]$
that assign a value to each pair (\texttt{rate\_of\_crimes}, \texttt{frequency\_of\_crimes}), where 
\texttt{rate\_of\_crimes} denotes the number of crimes in the hotspot divided  by the total of crimes and
 \texttt{frequency\_of\_crimes} is the temporal number of occurrences of the hotspot (computed for the binarized matrix~$H$) 
divided by total number of time slices. In fact, $f$ is simply a bilinear interpolation in the unit square
where $f(0,0)=0,\,f(0,1)=0.5,\,f(1,0)=0.7,\,f(1,1)=1$. With this distribution of values, we give more relevance
to hotspots where the number of crimes is larger.

\myparagraph{Selection:}
A hotspot selection filters the crimes in space and type. All the other views are recomputed to match the selected hotspot.

\myparagraph{Filtering:} 
Filtering the crimes using other views (\ie, space, time, or type) does not affect this view. 
If we want to recompute hotspots based on filtered data, for example, a particular crime type, we have to click the ``Hotspots'' button
after performing the data filtering. 

\subsection{Global Temporal View}
This view gives an overview of the number of crimes committed over the whole time period, relying on a line chart with a filled area between the data value and the base zero line (see \autoref{fig:teaser}(e)). 

\myparagraph{Time selection:} 
In this view, we can constraint the analysis at a particular time interval, which can be defined by brushing a rectangle on the \emph{Global Temporal View}. 
Only continuous time period can be selected. Next view will allow us to select multiple time intervals.
All views (except the hotspot that need to be recomputed)  are affected and automatically adjusted accordingly to the time selection. 

\begin{figure*}[t!]
    \centering
    \includegraphics[width=0.93\linewidth]{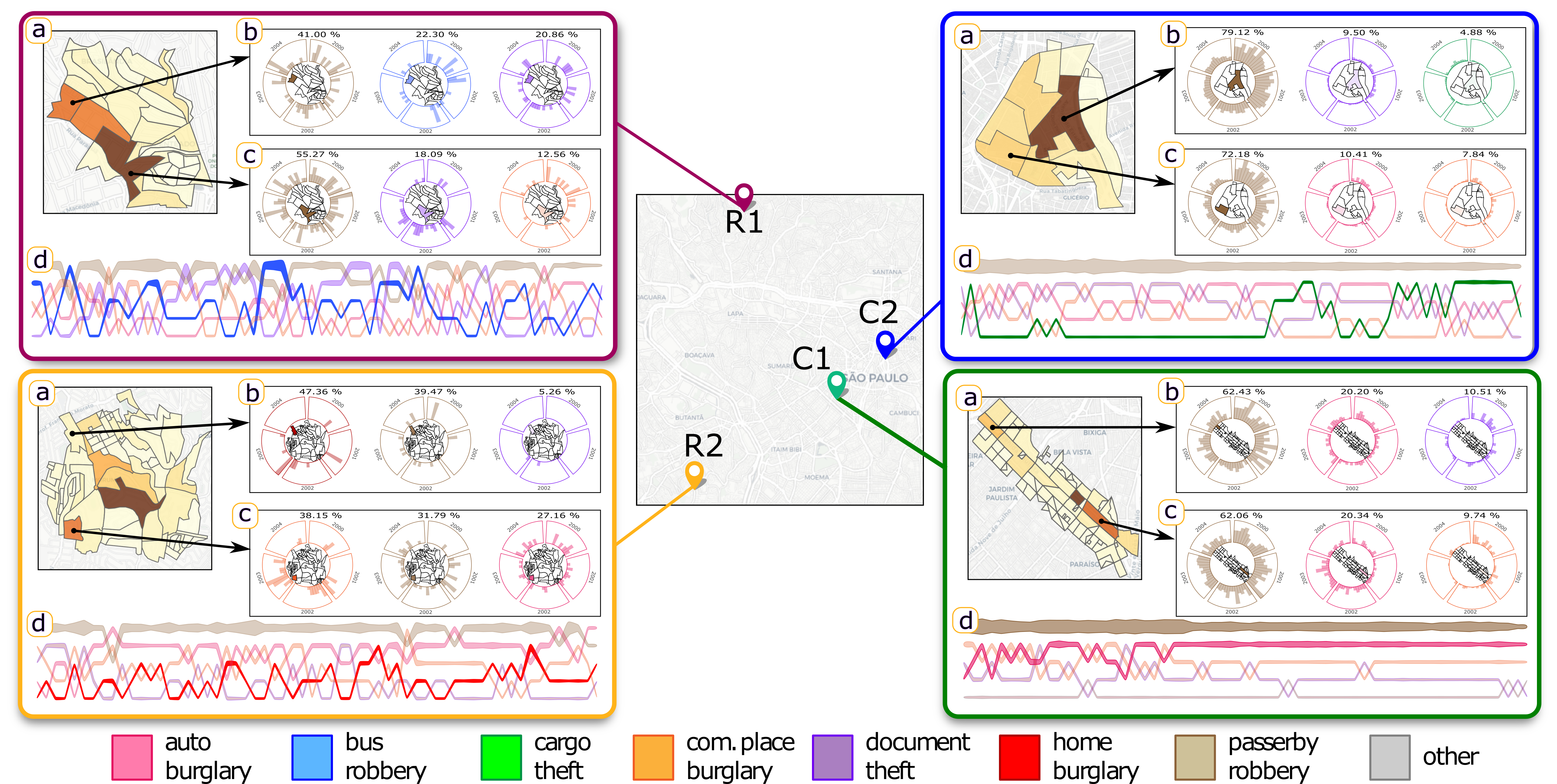}
    \vspace{-0.2cm}
    \caption{Summary of criminal activities and corresponding patterns in four different regions of S\~ao Paulo city. Crime patterns might change substantially
    among the regions and also along the time.}
    \label{fig:cs1}
    \vspace{-0.5cm}
\end{figure*}

\subsection{Cumulative Temporal View}
This view uses a bar chart to present the number of crimes accumulated by month, day and period of the day (see \autoref{fig:teaser}(d)).
In this view, we can see some patterns from non-continuous time intervals. This is also very useful to compare weekends or weekdays.

\myparagraph{Filtering:} 
When other views are used to filter the dataset, the filtered data is also overlaid on the global \emph{Cumulative Temporal View}, thus enabling a comparative analysis (see \autoref{fig:cs3}).

\subsection{Ranking Type View}
This view depicts three relevant pieces of information in a single metaphor: crime type evolution, crime types ranking, and a number of crimes in each time slice. 
As shown in~\autoref{fig:teaser}(f), each crime type is represented by a polyline. The vertical position, on each time step, encodes the relevance compared to others. Moreover, the line width is proportional to the number of crimes belonging to it.

\myparagraph{Filtering:} 
When a filter is activated in other views, the ranking view is recomputed using the filtered data. 

\subsection{Radial Type View}
In this view, we are using multiple bar charts with a radial layout. Each chart represents a different crime type, for instance, in \autoref{fig:teaser}(g) we have 5 crime types. In addition, the number on top of each chart shows the percentage for each crime type. 
Each chart is divided into sectors, where each sector is comprised of 12 bars depicting the months each year. 

\myparagraph{Crime Type selection:}
Clicking a chart filters the data to a specific crime type. In this way, users can focus their analysis on the  most crime-prevalent types. Selected crime types are represented by  a dashed border line.

\myparagraph{Time selection:}
We provide interactivity features on each chart to enable comparison among the same month on different years and same month across different crime types.

\myparagraph{Filtering:} 
When the dataset is filtered, each chart is recomputed to represented the filtered data.

\subsection{Filter Widget} 
This widget is comprised of a time and crime type histogram. For instance, \autoref{fig:teaser}(h) summarizes our data in four years (2000-2014) and five crime types. Moreover, we use this histograms to filter our data. Clicking a bar we can remove a year or a crime type. This filtering affects the whole interface.

\vspace{1.5\baselineskip}
\noindent Although most of the presented visual resources are not novel individually, many of them (such as \emph{hotspot view}, \emph{ranking type view}, and \emph{radial type view}) are nontrivial in the context of crime mapping. Even more important, the combination of all of them allows multiple analysis simultaneously, revealing interesting crime patterns as shown in the Case Studies (Sec.~\ref{sec:caseStudies}).

\section{Case Studies}
\label{sec:caseStudies}
This section presents three case studies that show the effectiveness of CrimAnalyzer in addressing the analytical
tasks presented in Sec.~\ref{sec:desiderata}. The first case study addresses analytical tasks T1, T2, and T3, while the second
is focused on hotspots analysis and related to T4 and T5. The third case study is aiming to make a parallel between criminal activity in S\~ao Paulo city and some crime related phenomena
reported in the literature and related to T3. 
In all case studies, except if explicitly stated, we used the robbery and burglary chunk of the dataset
as described in Sec~\ref{sec:data_set}, with a monthly discretization.

\subsection{Comparing Crime Patterns over the City (T1, T2, T3)}
\label{sec:cs1}
The goal of this case study is to analyze pattern of crimes in different regions of the city in order to 
understand how they change according to urban characteristics.
Moreover, we also investigate the temporal evolution of crime patterns in different regions.

To perform the study we selected four areas in S\~ao Paulo, two in the center of the city, denoted as C1 and C2 in 
\autoref{fig:cs1}, and two in residential areas, pointed as R1 and R2 in \autoref{fig:cs1}. 
C1 is a financial district, hosting the headquarter of important banks and financial institutions, while
C2 is a commercial area with many stores, an important metro terminal, and also several touristic places. 
Both C1 and C2 have a huge flow of people during the whole year.
Residential areas R1 and R2 differ in terms of the economic level of residents, R1 is a middle class neighborhood while
R2 is a richer area, with luxurious buildings and houses. 

\autoref{fig:cs1} bottom right depicts region C1, selected by drawing a polyline along the main avenue of the financial district (T1), and highlights the \emph{radial type view} (C1-c) of the three most prevalent crime types of two sites in C1 (indicated by the arrows). 
The \emph{ranking type view} (C1-d) on the bottom shows how the incidence of the five most frequent crimes varies
along the time.
Two crime types lead the ranking along the years (the beige and pinkish curves on top),
\texttt{passerby robbery} and \texttt{auto burglary}.
By analyzing the \emph{radial type view} (C1-b and C1-c) of the highlighted sites, one can notice that those two crime types are indeed the prevalent ones
in those regions (encoded by the color). Inspecting other sites by simply clicking on them on the map view, we concluded that 
\texttt{passerby robbery} and \texttt{auto burglary} are the prevalent crime types in almost all sites in C1.

Performing the same analysis in region C2 (top right in \autoref{fig:cs1}), which was selected by clicking and expanding the 
central site of the region (the brownish one), we observe a different behavior.
The \emph{ranking type view} (C2-d) shows that there is one crime type that has been grown over the years (green curve), \texttt{cargo theft}. 
Selecting \texttt{cargo theft} from the \emph{radial type view} in the CrimAnalyzer interface (\autoref{fig:teaser}(h)),
the \emph{map view} (\autoref{fig:teaser}(b)) reveals that \texttt{cargo theft} is not prevalent in the whole region,
but it is concentrated in just a few sites, being the dark brown site
in the center of the region. Notice that \texttt{cargo theft} became the third most common crime type in that region over time, 
being behind only of \texttt{passerby robbery} and \texttt{document theft}.
Other sites present a more uniform behavior, having 
\texttt{passerby robbery}, \texttt{auto burglary}, and \texttt{commercial establishment burglary} as the main crime types .

Moving from the city center to more residential areas, the analysis reveals a substantial change in crime patterns,
as one can observe on the left of \autoref{fig:cs1}, where the crime pattern in R1 and R2 is summarized. 
In the residential region R1, for example (top left in \autoref{fig:cs1}),
\texttt{passerby robbery} remains the most common crime type, followed by \texttt{document theft}. However, 
some sites in R1 have \texttt{bus robbery} (passengers and/or drivers of public bus service are robbed) as the second most common criminal activity (R1-d). 
The orange site pointed out by the top arrow is an example (R1-b).
Site-by-site crime pattern analysis is easy to perform with CrimAnalyzer in this case, since the number of sites is mild
and users need only to select the site on the map to make its crime pattern revealed.
The importance of \texttt{bus robbery} in R1 is easily noticed in the 
\emph{ranking type view} (R1-d) depicted on the bottom, where the blue
curve  (\texttt{bus robbery}) reaches high rank levels in several opportunities.

Similarly to what happen in C1, C2, and R1 (and also in most of the city), region R2 (bottom left in \autoref{fig:cs1}) does have
\texttt{passerby robbery} as the predominant crime type, what can clearly be seen from the \emph{ranking type view} (R2-d). 
However, crime patterns vary considerably among the sites and some of them do not even have \texttt{passerby robbery} as the prevalent crime, as the 
two highlighted sites, which have \texttt{passerby robbery} as second in importance (R2-c and R2-d).
Moreover, \texttt{home burglary} is the most relevant crime in one of those regions. In fact, home burglary is a relevant crime in R2 as a whole,
as indicated by the reddish curve in the \emph{ranking type view} (R2-d). Notice that \texttt{home burglary} has increased in importance over the years. 

\vspace{1\baselineskip}
The discussion above shows that the visual analytic functionalities implemented in CrimAnalyzer
are able to sort out analytical tasks T1, T2, and T3 in a simple, intuitive, and effective way. The flexibility to handle spatially complex neighborhoods
with different shapes allows users to scrutinize set of blocks as well as regions along avenues and streets (analytical task T1). 
The combination of the \emph{ranking type view} and the \emph{radial type view} allows users to understand 
crime pattern in each region and also in particular sites, making evident how crime patterns change around the city
and even from site to site in a particular region (analytical task T2), a task difficult to be performed with direct searches in the crime database. 
In particular, \emph{ranking type view} and \emph{radial type view} turn out quite effective in revealing 
the temporal behavior of crime patterns, making clear that patterns have changed along the years (analytical task T3). 
With non-visual resources, this analysis could be an arduous process, demanding the implementation of multiple filters 
and sophisticated numerical and computational tools.

\subsection{Hotspot Analysis and Cargo Theft (T1, T4, T5)}
\label{sec:cs2}
\begin{figure}[t!]
	\centering
	\includegraphics[width=0.99\linewidth]{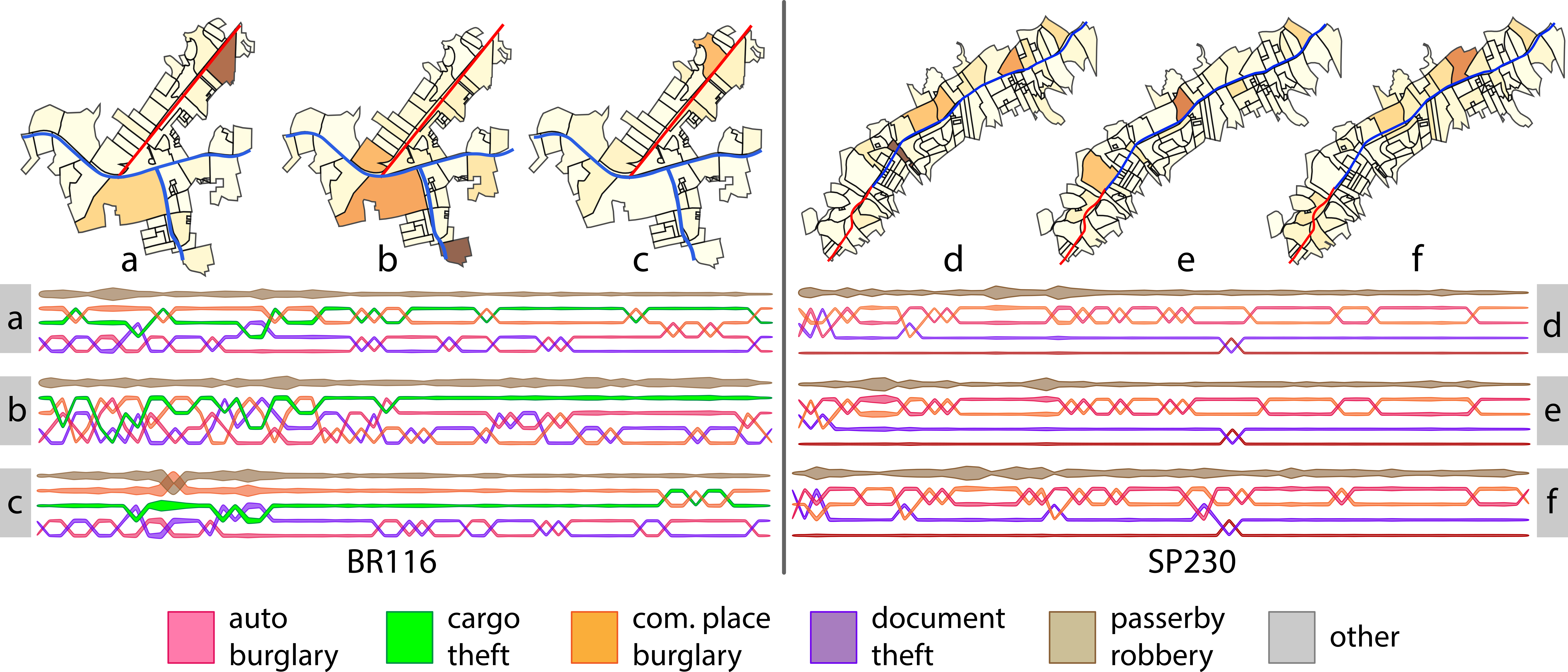}
  \vspace*{-0.3cm}
  \caption{Hotspots around BR116 and SP230 considering all crime types.}
	\label{fig:cs2}
  \vspace*{-0.3cm}
\end{figure}

This case study has been driven by the domain experts, they were interested on
a particular type of crime, \texttt{cargo theft}. Although \texttt{cargo theft}
does not figure among the most prominent crime types in S\~ao Paulo city, it is
of great interest due to its spatial characteristic, the high values involved,
and the engagement of violent gangs in this type of criminal activity. It is
well known that robbery (or theft) of high valuable cargo commodities tends to
happen close to the main highways connecting S\~ao Paulo to other regions of
Brazil. Therefore, domain experts focused their analysis in two important
highways, SP230, which connects S\~ao Paulo to states in the South of Brazil,
and BR116, which connects S\~ao Paulo to Rio de Janeiro.

In order to perform their analysis, domain experts relied on the polyline
selection tool to select a considerable number of sites along the highways and
avenues that connect the city to the highways. The number of regions involved in these
analysis renders a site-by-site investigation tedious, making hotspots a better
alternative. \autoref{fig:cs2} shows three \emph{hotspots} obtained from the
regions selected along BR116 and SP230 and nearby avenues. The highways are highlighted in red and
the nearby avenues in blue in the hotspot maps depicted in \autoref{fig:cs2}.
The \emph{ranking type view} reveals the crime patterns in each
hotspot (considering only the 5 most relevant crime types). Notice that in
BR116, \texttt{cargo theft} figures among the most relevant crimes (green lines), becoming the second most relevant crime in several
moments. 
In SP230, though, \texttt{cargo theft} is not
predominant, not appearing among the five most relevant crimes in the
\emph{ranking type view} in any hotspot. In SP230, the predominant pattern
is \texttt{passersby robbery}, \texttt{vehicle burglary}, and
\texttt{commercial establishment burglary}. CrimAnalyzer makes clear 
which sites are relevant in each hotspot, their crime patterns, and how crime
patterns evolve, thus properly tackling analytical tasks T4 and T5.

\begin{figure}[t!]
	\centering
	\subfigure[BR116]{\label{fig:cs2_dutra}\includegraphics[width=0.98\linewidth]{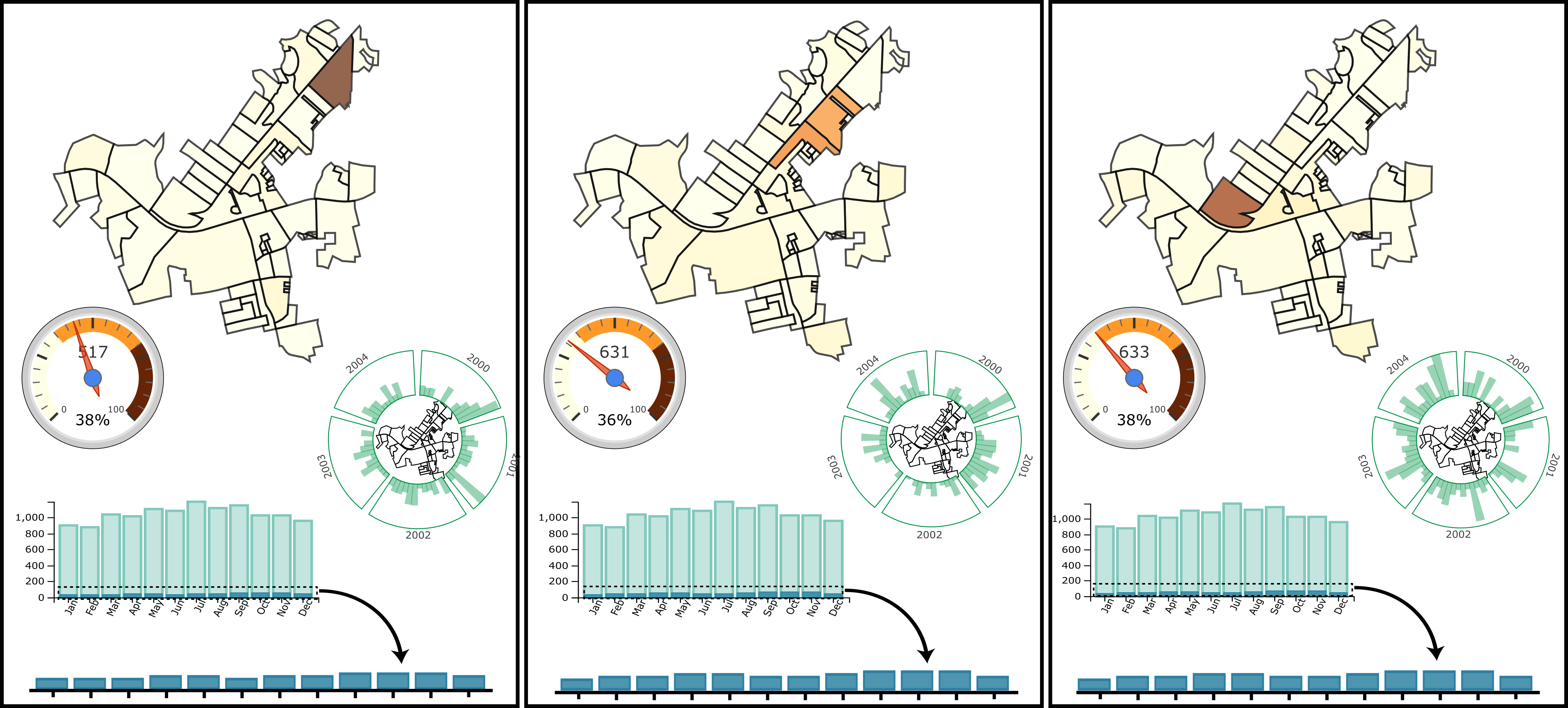}} \\ 
	\vspace*{-0.3cm}    \subfigure[SP230]{\label{fig:cs2_regis}\includegraphics[width=0.98\linewidth]{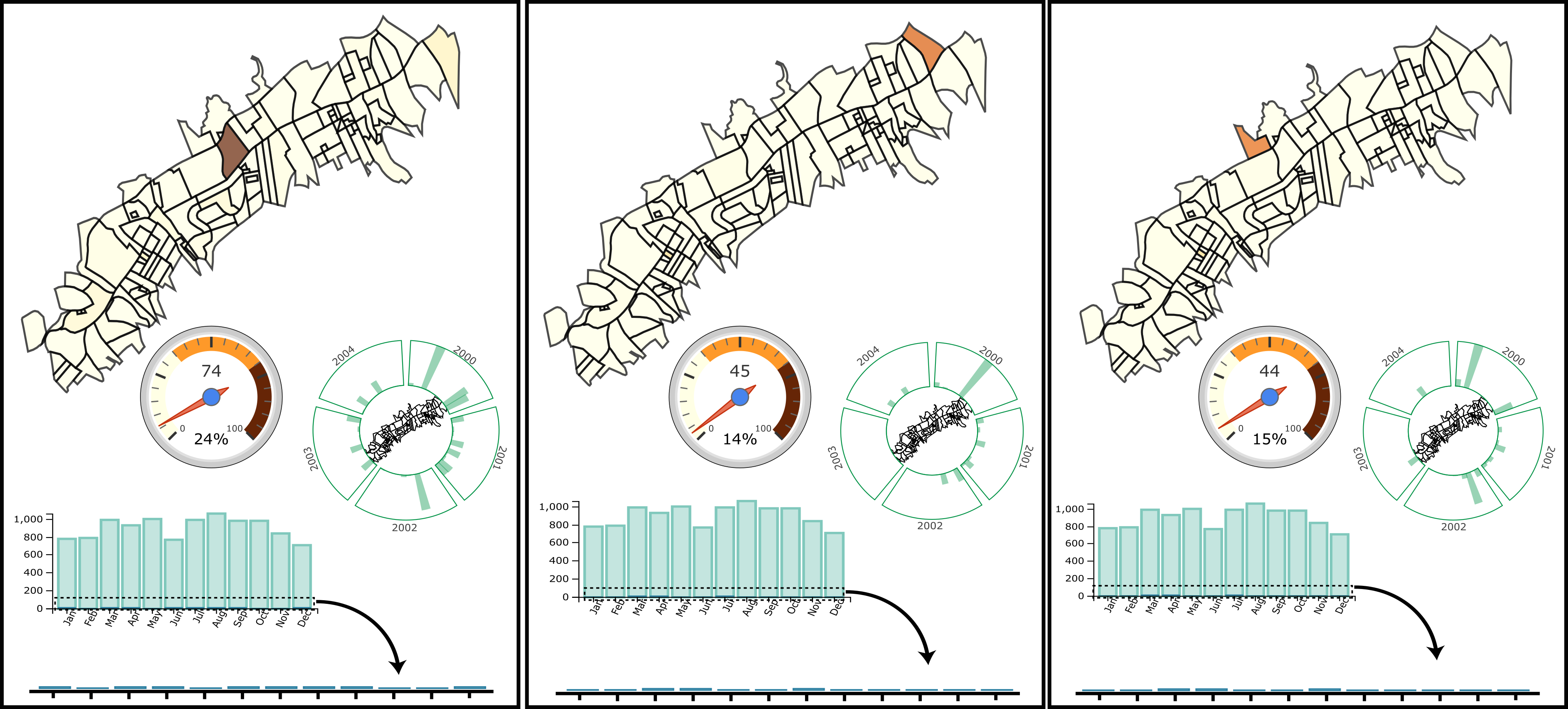}} 
	\vspace*{-0.3cm}
	\caption{\texttt{Cargo theft} hotspots along two important highways, BR116 and SP230. BR116 presents a much larger and more
		frequent number of \emph{cargo theft} than SP230.}
	\label{fig:cs2_regis_dutra}
	\vspace*{-0.4cm}
\end{figure}

However, the experts are particularly interested in \texttt{cargo theft}. To
center the analysis in a single crime type users only need to select that crime
type in the \emph{radial type view}, filtering the data such that
hotspots and all the views are updated to depict only information related to
the selected crime type. \autoref{fig:cs2_regis_dutra} shows the hotspots
associated to \texttt{cargo theft} only.
The \emph{gauge widgets} show that the number of \texttt{cargo theft} in the
BR116 is one order of magnitude larger than in SP230, also presenting a higher
rate of occurrence. 
The temporal evolution (\emph{radial type view}) on the center right of each \emph{grid} shows the temporal behavior of \texttt{cargo theft} 
in each hotspot. It is clear that the number of \texttt{cargo theft} in SP230 has
lessened over the years, while in BR116 no reduction is observed. The histograms below the gauge widget show the intensity of \texttt{cargo theft} (the short dark
bars) in each month, comparing them against the total number of crimes in the region.

The CrimAnalyzer viewing tools also make clear that, in BR116, \texttt{cargo theft} takes
place mainly along the highway (red curves in BR116 maps in \autoref{fig:cs2}), 
while in SP230 the relevant sites of each
hotspot are located in the avenue that connects the highway to the city (blue curves in SP230 maps in \autoref{fig:cs2}).
Domain experts considered this an important finding, mainly to make public security
policies more efficient. 
Another interesting aspect pointed out by the experts
is the capability of revealing hotspots associated with sparse criminal
activities, as the one depicted in \autoref{fig:cs2_regis} (see the spikes in the radial view).
Sparse hotspots are relevant and deserve to be investigated.
Notice that these finds could hardly be made without the visual resources enables by CrimAnalyzer.

\subsection{Seasonality and the Temporal Element of Crime (T3)}
This case study corroborates if some criminal behaviors described and validated 
in previous works also take place in S\~ao Paulo.

\begin{figure}[t!]
    \centering
        \subfigure[\small Commercial district 1.]{\label{fig:cs3_itaim}\includegraphics[width=0.31\linewidth]{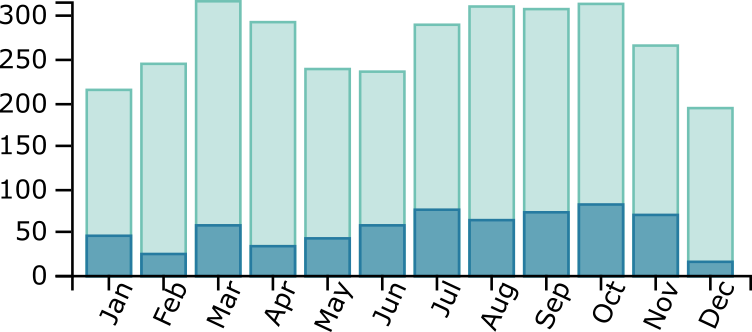}} 
        \subfigure[\small Commercial district 2.]{\label{fig:cs3_vima}\includegraphics[width=0.31\linewidth]{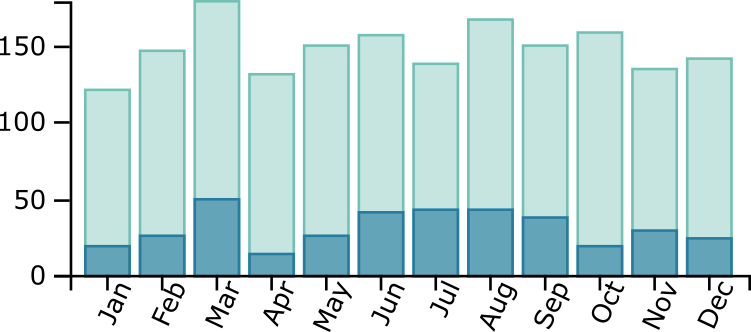}} 
        \subfigure[\small Commercial district 3.]{\label{fig:cs3_liberdade}\includegraphics[width=0.31\linewidth]{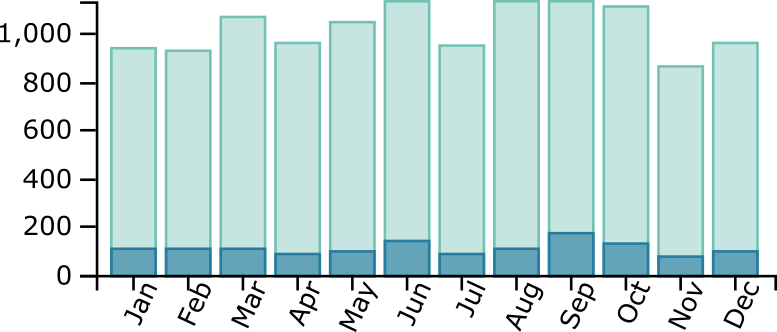}} \\ 
        \subfigure[Commercial street 1.]{\label{fig:cs3_eusma}\includegraphics[width=0.31\linewidth]{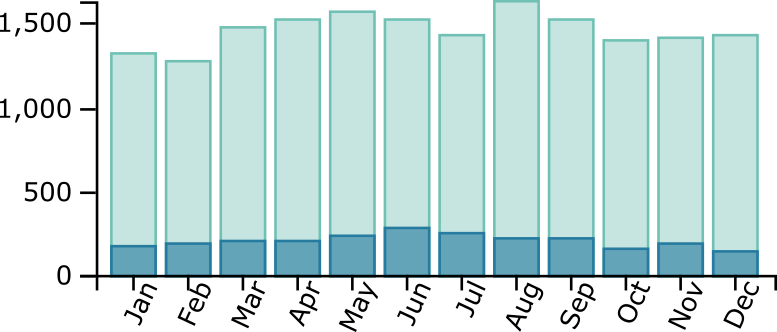}} 
        \subfigure[Commercial street 2.]{\label{fig:cs3_teosam}\includegraphics[width=0.31\linewidth]{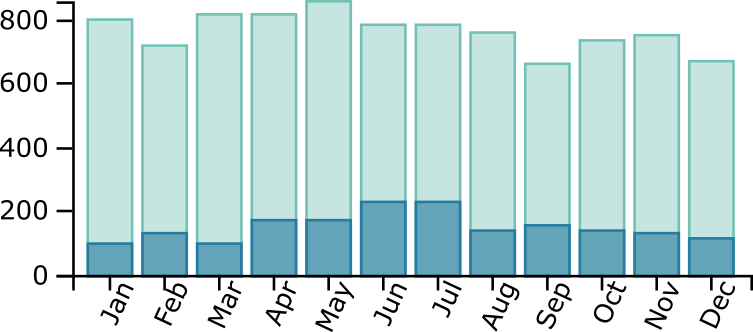}} 
        \subfigure[Commercial street 3.]{\label{fig:cs3_25}\includegraphics[width=0.31\linewidth]{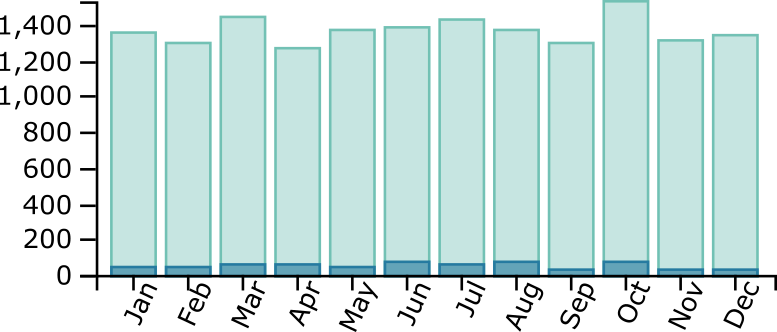}} 
    \vspace*{-0.3cm}
    \caption{Commercial establishment burglary tends to increase during the winter (winter in South America goes from mid June to mid September).}
    \label{fig:cs3}
    \vspace*{-0.2cm}
\end{figure}

\myparagraph{Seasonality} An important aspect related to criminal activities is
seasonality. There is a number of studies in the literature that support the
hypothesis that certain crime types are seasonal while others are not. For
instance, van Koppen and Jansen~\cite{van1999time} argue that, in Netherlands,
commercial establishment burglary (robbery) increases during the winter due to
the increased number of dark hours during the day. In South America, winter
usually starts in mid June and last until mid September, during this period,
especially in July and August, the number of dark hours is higher than in the rest
of the year. An interesting question related to analytical task T3 is whether
the findings of van Koppen and Jansen is valid in S\~ao Paulo city. To
search an answer, we relied on CrimAnalyzer to explore six major commercial
areas in S\~ao Paulo city, three commercial districts
and three popular commercial streets. 
\autoref{fig:cs3} shows the \emph{cumulative temporal view} 
of each of the analyzed regions.
The overlaid darker histograms correspond to the number of \texttt{commercial establishment burglary} and \texttt{robbery}
in each month. The overlaid histogram is generated by simply selecting \texttt{commercial establishment burglary}
in the \emph{temporal type view}.

From \autoref{fig:cs3} one clearly sees that five out of six regions present an
increase in the number of \texttt{commercial establishment burglary} and
\texttt{robbery} during the winter (a-e), thus supporting the
findings of van Koppen and Jansen. Although we can not claim with certainty
that the hypothesis is true, the analysis enabled by CrimAnalyzer provides
evidences about the seasonality of this type of crime, thus helping to
answer one of the questions associated to analytical task~T3.

\myparagraph{Near Repeat Victimization} Near repeat victimization theory claims that when a home is burgled, the risk of 
recidivism is not only higher for the targeted home, but also for the nearby homes, with risk period that seems to decay 
after some weeks or months~\cite{polvi1991time}. The near repeat victimization theory has found evidences of its veracity 
in a number of countries, but we could find no report about it in S\~ao Paulo. 

\begin{figure}[t!]
	\centering
	\includegraphics[width=0.95\linewidth]{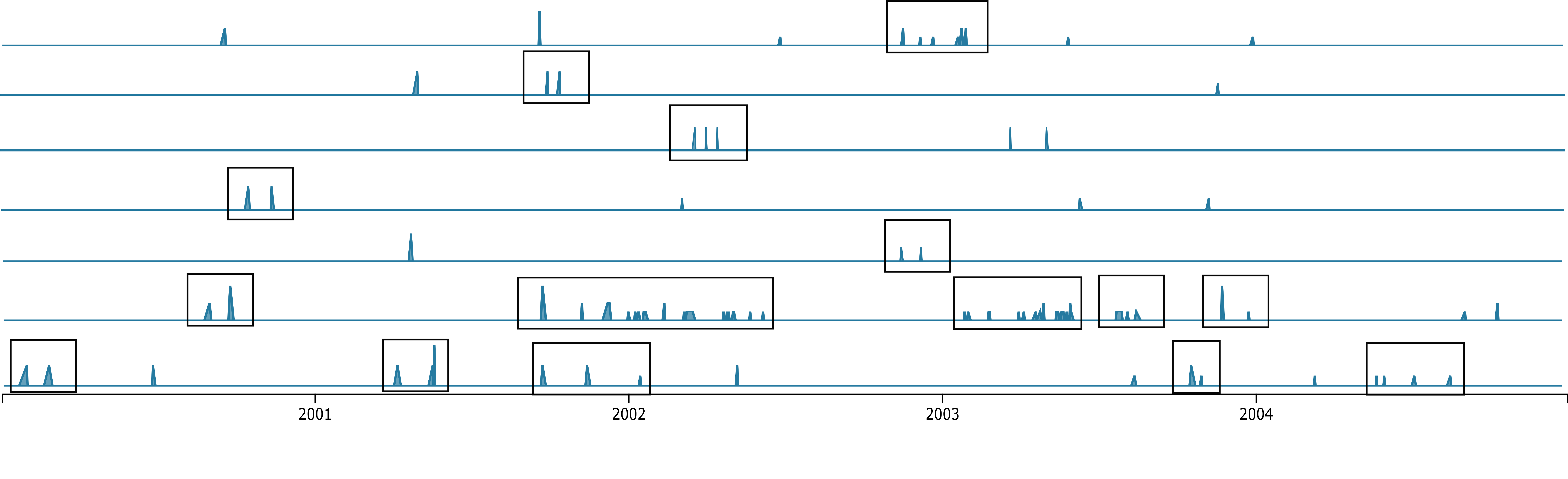} 
	\vspace*{-0.2cm}
	\caption{The \emph{Near Repeat Victimization} phenomena. When a home is burgled, the risk of
		recidivism in a short period of time is not only higher for the targeted home, but also for the nearby homes.}
	\label{fig:cs3_home}
	\vspace*{-0.1cm}
\end{figure}

Using CrimAnalyzer, we scrutinized two regions in S\~ao Paulo city where \texttt{home burglary} is a recurrent crime, 
including region R2 discussed in the case study presented in Sec.~\ref{sec:cs1}. 
\autoref{fig:cs3_home} shows the time series, in a daily temporal scale, of seven sites in the analyzed regions, 
which varies in terms of the frequency of crimes and the number of \texttt{home burglary}. 
The boxed spikes point \texttt{home burglary} events that occur less than thirty days apart from each other. 
Notice that even in sites where \texttt{home burglary} is really occasional (rows 2 to 5 in \autoref{fig:cs3_home}), 
The time series can be visualized in CrimAnalyzer by switch the time discretization to the scale of days and select
the region of interest in the map view.

Although the validation of \emph{seasonality} and \emph{Near Repeat Victimization} can be done with non-visual methods, with the functionalities implemented in CrimAnalyzer, this validation could be direct and simple.
As far as we know, these phenomena have not yet been reported in Brazil, being this an important side contribution of this work. 

\vspace*{-0.1cm}

\section{Evaluation from the Experts}

After playing with CrimAnalyzer and run a number of experiments, including the case study reported in Sec.~\ref{sec:cs2},
domain experts have sent us the following feedback.

``\emph{Despite its limitations, CrimAnalyzer has allowed us to better understanding challenges not yet elucidated by conventional crime analysis tools.
First, by using solid mathematical and computational resources to reveal geo-referenced criminal activities, CrimAnalyzer incites 
the search for plausible explanations for the observed criminal patterns, what would be impossible with conventional analysis. 
Second, CrimAnalyzer motivates reflection about the relationship among the different crime types and about topological, directional, and
relational connections that might affect the number of crimes in specific locations and time intervals.
Third, an analytical tool that enables the analysis of crimes in specific locations leads to thinking the city in its complexity
and, at the same time, guides the investigation of urban characteristics (administrative, demographic, physical, and social) 
and their interaction from which the observed local patterns result. 
Fourth, CrimAnalyzer uncovers the heterogeneity of the city as to its urban infrastructure, the differences among commercial, financial, and 
residential areas, the flow of people, public and private transportation, as well as the need for improvements, not only in
terms of policing in specific locations and according to the type of crimes, but also, and mainly, in terms of tools to 
assist criminal investigation towards reducing the high rates of impunity. 
Finally, in contrast to more simplistic statistical methodology, the deterministic approach for hotspot identification turns out 
fundamental to emphasize the dynamics of spatio-temporal processes and to capture typical social manifestations such as 
crimes.}'' 

The experts were quite enthusiastic about CrimAnalyzer, as it allowed them to understand and create new hypotheses about the behavior of the crime type \texttt{cargo theft}. Specifically, one mentioned ``\emph{With the use of this tool, we can put in the hands of the decision takers clear evidence of how public policies should be defined with respect to car thefts}".

\section{Discussion and Limitations}
CrimAnalyzer was developed in close cooperation with domain experts. The current version satisfies their the requirements, 
however, some limitations and future work leaped out of this collaboration.

\myparagraph{NMF stability.}
Our approach for identifying hotspots is not stable, this is because the Non-Negative Matrix Factorization technique depends on the initial conditions of the optimization procedure. To counteract this effect, some implementations, like the one we are using in the presented system, 
enables us to run the method a number of times, keeping the solution with the smallest error. 
Although the results got quite stable after enabling the multiple run alternative, a more robust 
approach could be sought to mitigate possible effects.

\myparagraph{Space Discretization.}
The space discretization used in CrimAnalyzer is the census units in S\~ao Paulo city, we adopted this measure because our collaborators had interest in seeing the analysis in this level of detail. However, we are aware of the modifiable areal unit problem (MAUP), census units do not represent ``natural units'' of analysis and
the result of certain analysis can change by modifying the aggregation unit~\cite{GIS_Book:2005}. An immediate future work would be to extend and make more flexible our space discretization. In this way, we should be able to apply our tool in other contexts.

\myparagraph{Multiple data sources.}
Crime events by their own rarely tell the whole story. Additional data that can be used to enhance the understanding of the crime layer.
For example, the presence of bars and pubs, distance to parks, vacant land and buildings, weather, among other information might have 
a relation with certain criminal activities. 
Given the increasing number of initiatives to make data publicly available, we are considering to combine that information to further understanding crimes in urban areas. An interesting mathematical tool in this context is tensor decomposition, 
a generalization of matrix decomposition able to extract patterns from multiple data sources. 
Developing visual analytical tools to map tensor decomposition information into visual content is an important problem~\cite{ballester2019tthresh} 
that has barely been approached in the context of crime analysis.

\myparagraph{Global vs Local approach.} 
CrimAnalizer uses a local-based approach to explore and analyze crime patterns. Even though this was a requirement from the domain experts, and we agree that it was the correct approach to this problem, mainly because domain experts have prior knowledge and hypothesis regarding crime behaviors in particular locations, in some of our interviews with domain experts we discussed the option of having a global-based technique that might process the whole space and propose interesting locations to be explored. This alternative was accepted by the experts but as a complementary technique. 
As future work, we are also interested in tackling this problem from both perspectives (global and local).

\myparagraph{Multiple cities and different scenarios}
Finally, we intend to apply and validate our system in other cities and countries. Currently, we are in the process of collecting crime data from multiple locations, and in a short time, we expect to release the system to analyze multiples cities in Brazil. 
In addition, our approach can be extended to other scenarios than crime analysis.
For instance, one can use the system to analyze the dynamics of traffic-accidents in particular locations of the city,
making possible to uncover how the number of car-car crashes, car-bus crashes, run overs, etc. evolve over time.

\vspace*{-0.1cm}

\section{Conclusion}
We introduced a visualization assisted analytic tool to support the analysis of crimes in local regions. 
We developed CrimAnalyzer in close collaboration with domain experts and translated their analytical into the visualization system. 
We also propose a technique based on NMF to identify hotspots. Our system was validated by qualitative and quantitative comparisons, and case studies using real data and with feedback from the domain experts. Moreover, we verified two crime behavior facts (\ie, \emph{seasonality} and \emph{near repeat victimization}) using S\~ao Paulo crime data.

\acknowledgments{
This work was supported by CNPq-Brazil (grants \#302643/2013-3 and  \#301642/2017-63), CAPES-Brazil, and S{\~a}o 
Paulo Research Foundation (FAPESP) - Brazil. The views S{\~a}oPaulo Research Foundation (FAPESP) - Brazil (grant \#2016/04391-2 and \#2017/05416-1). The views
expressed are those of the authors and do not reflect the official
policy or position of the S{\~a}o Paulo Research
Foundation. We also thanks Intel for making available part of the computational resources we use
in the development of this work.
}

\balance

\bibliographystyle{abbrv-doi}

\bibliography{refs}
\end{document}